\documentclass[fleqn,usenatbib]{mnras}

\usepackage{newtxtext,newtxmath}

\usepackage[T1]{fontenc}
\usepackage{ae,aecompl}

\usepackage{graphicx}	
\usepackage{amsmath}	
\usepackage{amssymb}	

\usepackage{float}
\usepackage[figure,figure*]{hypcap}
\usepackage{subfigure}
\usepackage{soul}

\usepackage[dvipsnames]{xcolor}

\title[Deep-learning real/bogus classifier for ZTF]{Real-bogus classification for the Zwicky Transient Facility using deep learning}

\author[D. A. Duev et al.]{Dmitry A. Duev,$^{1}$\thanks{E-mail: duev@caltech.edu (DAD)}
    Ashish Mahabal,$^{1}$
    Frank J. Masci,$^{2}$
    Matthew J. Graham,$^{1}$
    \newauthor
    Ben Rusholme,$^{2}$
    Richard Walters,$^{1}$
    Ishani Karmarkar,$^{3}$
    Sara Frederick,$^{4}$
    \newauthor
    Mansi M. Kasliwal,$^{1}$
    Umaa Rebbapragada,$^{5}$
    Charlotte Ward,$^{4}$
\\
$^{1}$Division of Physics, Mathematics, and Astronomy, California Institute of Technology, Pasadena, CA 91125, USA\\
$^{2}$IPAC, California Institute of Technology, MS 100-22, Pasadena, CA 91125, USA\\
$^{3}$Division of Engineering and Applied Science, California Institute of Technology, Pasadena, CA 91125, USA\\
$^{4}$Department of Astronomy, University of Maryland, College Park, MD 20742, USA\\
$^{5}$Jet Propulsion Laboratory, California Institute of Technology, Pasadena, CA 91109, USA\\
}

\date{Accepted XXX. Received YYY; in original form ZZZ}

\pubyear{2019}

\begin{document}
\label{firstpage}
\pagerange{\pageref{firstpage}--\pageref{lastpage}}
\maketitle

\begin{abstract}
Efficient automated detection of flux-transient, reoccurring flux-variable, and moving objects is increasingly important for large-scale astronomical surveys. We present \texttt{braai}, a convolutional-neural-network, deep-learning real/bogus classifier designed to separate genuine astrophysical events and objects from false positive, or bogus, detections in the data of the Zwicky Transient Facility (ZTF), a new robotic time-domain survey currently in operation at the Palomar Observatory in California, USA.
\texttt{Braai} demonstrates a state-of-the-art performance as quantified by its low false negative and false positive rates. 
We describe the open-source software tools used internally at Caltech to archive and access ZTF's alerts and light curves (\texttt{Kowalski}), and to label the data (\texttt{Zwickyverse}).
We also report the initial results of the classifier deployment on the Edge Tensor Processing Units (TPUs) that show comparable performance in terms of accuracy, but in a much more (cost-) efficient manner, which has significant implications for current and future surveys. 
\end{abstract}

\begin{keywords}
methods: data analysis -- surveys
\end{keywords}



\section{Introduction and context}

Astronomical sky surveys observe a plethora of transient events in the dynamic sky originating from a wide range of astrophysical objects and processes. Detection of such events can be performed in the catalog domain (e.g., CRTS survey; \citet{Drake2009}) and/or in the image domain
(e.g., PTF survey; \citet{Law:09:PTFOverview}). In the latter case, an epochal image of a patch of the sky is compared to a reference image, which is usually achieved by means of image subtraction. In the process, time-dependent characteristics of the images such as the point spread functions (PSF) and depth are matched. There are multiple factors that may lead to false positive, or bogus, detections in the resulting subtracted images:

\begin{itemize}
\item Unmodeled differences between the images that are present even in the idealized situation of noise absence, e.g. radiation hits, optical ghosts, persistent charge, and imperfections in flat-fielding.

\item Noise and, most importantly, its unmodeled components, e.g. registration errors, source noise errors, and incorrect estimates of the noise components.
\end{itemize}

Current and future large-scale surveys have the ability to detect millions of subtraction residuals a night, manifesting the need for automated separation of genuine astrophysical events from bogus detections. Both the real and bogus events may be caused by a wide variety of phenomena, some of which are very hard to model. For example, there is no proper statistical model for radiation hits and optical ghosts. Therefore, an explicit programmatic solution to the problem is difficult and it is most efficient to apply machine learning (ML) methods to extract the relevant patterns from the data themselves.

The real/bogus (RB) ML classifiers score individual sources on a scale from 0.0 (bogus) to 1.0 (real). RB classifiers were first introduced by \citet{2007ApJ...665.1246B} for the Nearby Supernova Factory \citep{2002SPIE.4836...61A}, and have been adopted by other time domain surveys including the Palomar Transient Factory \citep[PTF;][]{bloom_towards_2008} and the Intermediate Palomar Transient Factory \citep[iPTF;][]{brink_using_2012,wozniak_automated_2013,rebbapragada_time-domain_2015}, the Dark Energy Survey \citep{goldstein2015des}, Pan-STARRS \citep{wright2015panstarrs}, and HiTS \citep{2017ApJ...836...97C, 2018arXiv180803626R}.

\subsection*{The Zwicky Transient Facility}

The Zwicky Transient Facility (ZTF) is a new robotic time-domain sky survey capable of visiting the entire visible sky north of $-30^\circ$ declination every night. ZTF observes the sky in the $g$, $r$, and $i$ bands at different cadences depending on the scientific program and sky region \citep{2019PASP..131a8002B, 2019PASP..131g8001G}. The new 576 megapixel camera with a 47 deg$^2$ field of view, installed on the Samuel Oschin 48-inch (1.2-m) Schmidt Telescope, can scan more than 3750 deg$^2$ per hour, to a $5\sigma$ detection limit of 20.7 mag in the $r$ band with a 30-second exposure during new moon \citep{Dekany2019, 2019PASP..131a8003M}.

The raw data are transferred to the Infrared Processing and Analysis Center (IPAC) at the California Institute of Technology (Caltech) and processed in real time. The ZTF Science Data System (ZSDS) housed at IPAC consists of the data processing pipelines, data archives, infrastructure for long-term curation, and the services for data retrieval and visualization. For the detailed description of the ZSDS please refer to \citet{2019PASP..131a8003M}.

The part of the ZSDS responsible for the transient event detection and extraction, first ``properly'' subtracts a reference (template) image from a calibrated science exposure image. In summary, this step involves using a subset of sources from the input reference and epochal (science) PSF-fit photometry catalogs to match the photometric throughputs of the corresponding images; resamples and interpolates the reference image onto the science image using \texttt{SWarp} \citep{2010ascl.soft10068B}; masks all bad pixels propagated from the science and reference images; computes a smoothly varying differential background image and subtracts this from the science image. Pixel-uncertainty images and PSFs for the science and reference images are then generated. PSF matching and image differencing are then performed using the ZOGY algorithm \citep{2016ApJ...830...27Z}.

If the resulting difference image is of sufficient quality, the pipeline then detects events from the point-source match-filtered S/N images where detection is performed on both the positive (science minus reference) and negative (reference minus science)\footnote{The negative images are simply $-1\times$ the positive images generated by a single run of the ZOGY software.} images. Events are extracted with both aperture and PSF-fit photometry, and additional source features are computed. The events are then lightly filtered to remove obvious false positives and image cutouts are generated \citep{2019PASP..131a8003M}. 

Events may have been triggered from a flux-transient, a reoccurring flux-variable, or a moving object. The metadata and contextual information including the cutouts are put into ``alert packets'' that are further picked up by the ZTF Alert Distribution System (ZADS). On a typical night, the number of detected events ranges from $10^5$ -- $10^6$.

The RB classifier initially employed by ZTF heavily relied on the PTF/iPTF legacy and was built using the random forest (RF) algorithm. To make a prediction, it used the source features extracted from the science and subtracted image cutouts centered on the candidate, supplemented with other measurements taken from the science, subtracted, and reference images. Please refer to \citet{2019PASP..131c8002M} and Rebbapragada et al. (in prep.) for the details on the RF RB classifier.

In this paper, we present \texttt{braai}\footnote{Bogus-Real Adversarial Artificial Intelligence}, a new cutout-image-based RB classifier built for ZTF using deep learning (DL) that demonstrates a state-of-the-art performance, superior compared to the original RF classifier.\footnote{We note that the RF RB scores given in this work are from the alert packets and come from multiple versions of the classifier.} Additionally, we describe the open-source software tools used internally at Caltech to archive and access ZTF's alerts and light curves (\texttt{Kowalski}), and to label the data (\texttt{Zwickyverse}).

\section{braai: a Deep Learning framework for real-bogus classification}

Deep learning is a subset of ML that employs artificial many-layer neural networks \citep{McCulloch1943}. DL systems are able to discover, in a highly automated manner, efficient representations of the data, simplifying the task of finding the meaningful sought-after patterns in them.

\subsection{Data set}

DL systems are able to learn even very complicated, highly non-linear mappings between the input and output spaces reaching near-optimal performance. The challenge is to construct a large, labelled, representative data set for the network training. In case of the RB classification, the training set must reflect the possible variations across different filters, sky location, CCDs, as well as cross-talk.


In this work, we used a number of sources for data collection. The ZADS distributes the alert packets in the Apache Avro$^{TM}$ format\footnote{\url{https://avro.apache.org}} through the ZSDS Kafka\footnote{\url{https://kafka.apache.org}} cluster at IPAC \citep{2019PASP..131a8003M}. 

\begin{figure*}
  \centering
  \includegraphics[width=0.95\textwidth]{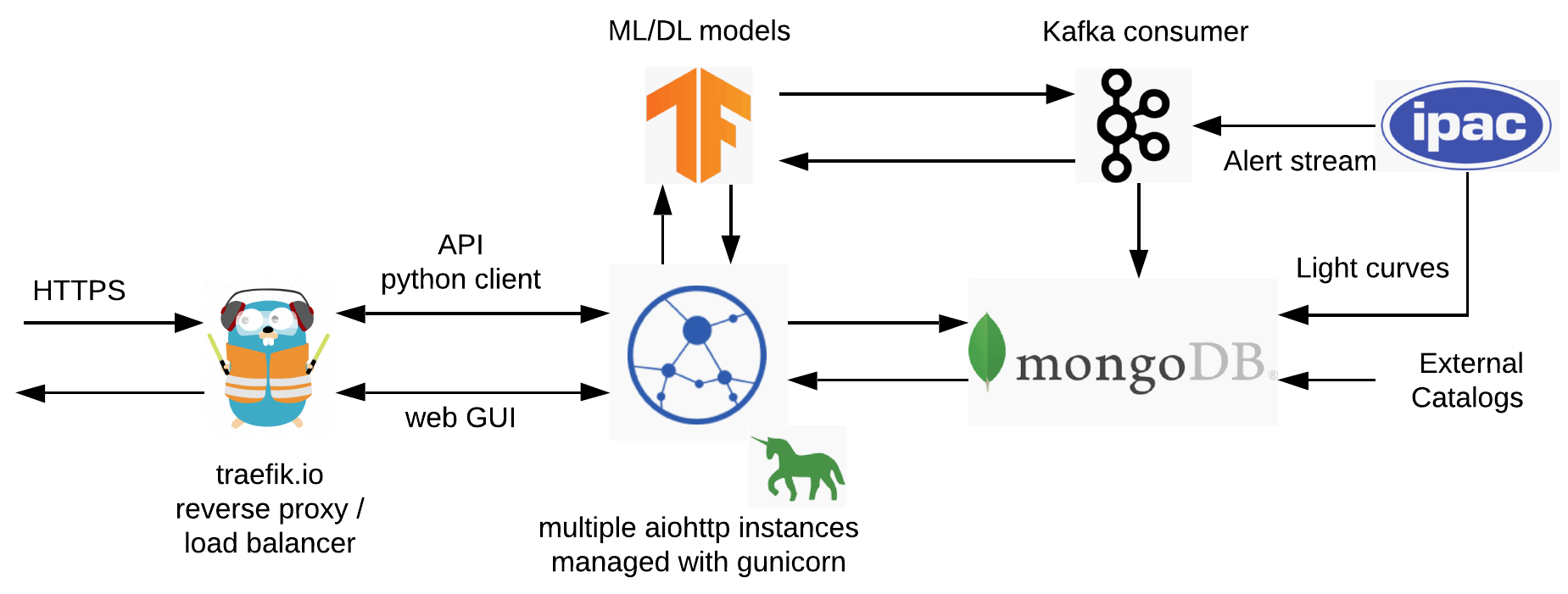}
    \caption{Architecture of \texttt{Kowalski}. See Section \ref{sec:kowalski} for a detailed description.}
    \label{fig:Kowalski.architecture}
\end{figure*}

Internally at Caltech, the alert stream is consumed by \texttt{Kowalski}\footnote{\url{https://github.com/dmitryduev/kowalski}}, an open-source system primarily used to archive and access ZTF's alerts and light curves (see Section \ref{sec:kowalski}). We queried \texttt{Kowalski} to gather samples of data representing the vast diversity of ZTF's alert parameter space.

Another internal consumer, the GROWTH Marshal \citep{Kasliwal_2019}, is used to analyze and coordinate follow-up of the sources discovered in the ZTF alert stream through programmatic filtering and human vetting. The GROWTH marshal served as the primary source of pre-labelled (mostly transient) events.

A large number of the bogus examples was collected at the start of the survey for the RF classifier since the initial focus was to filter out the majority of the typical artifacts present in the alert stream such as those caused by the bright stars.

Finally, a small chunk of pre-labelled data came from the Zooniverse Citizen Science platform\footnote{\url{https://www.zooniverse.org}}, where we set up a dedicated project (see \url{https://www.zooniverse.org/projects/rswcit/zwickys-quirky-transients}). These data were used for testing purposes (see Section \ref{sec:test_sets}).

\subsubsection{Kowalski}\label{sec:kowalski}

We developed \texttt{Kowalski} for the primary task of supporting the time-domain astronomy efforts with ZTF. Concretely, it solves the problem of efficient storage and access (both programmatic and GUI-based) through a standardized application programming interface (API) to both ZTF's alert/light-curve data and external catalogs.

\texttt{Kowalski}'s architecture is shown in Fig. \ref{fig:Kowalski.architecture}. The back-end is powered by a non-relational (NoSQL) database \texttt{MongoDB}\footnote{\url{https://mongodb.com}} with an API layer on top of it managing the incoming and outgoing traffic/data streams. We based the choice of \texttt{MongoDB} as the workhorse on the following reasons:\\

\begin{itemize}
\item Individual entries are stored as binary \texttt{JSON} (\texttt{BSON}) ``documents'' in ``collections''. This naturally maps to the format of the alert \texttt{AVRO} packets. The light curve data are stored per source, thus significantly reducing the number of necessary read operations when accessing the data.

\item Collections are implemented as B-trees, which guarantees $\sim log(N)$ execution times for the standard database CRUD (create, read, update and delete) operations, where $N$ is the number of documents in the collection.

\item Collections support multiple, potentially compound indexes and associated (frequently, in-memory) ``covered'' queries, providing efficient access to the most-in-demand data.

\item Being a NoSQL database, \texttt{MongoDB} does not enforce any schema by default meaning no downtime in case of an alert packet schema change.

\item Built-in \texttt{GeoJSON} support with 2D indexes on the sphere allowing efficient, potentially complicated positional queries.

\item Built-in support for horizontal scaling through sharding\footnote{A type of database partitioning that separates very large databases the into smaller, faster, more easily managed parts called data shards.}.

\end{itemize}

The API layer is built using a \texttt{python} asynchronous web framework \texttt{aiohttp}\footnote{\url{https://docs.aiohttp.org}}. Authorization is performed using the JSON web tokens. The standard \texttt{python} async event loop (with futures scheduling) serves as a simple, fast, and robust job queue. Both web-based graphical user interface (GUI) and a programmatic \texttt{python} client are available to interact with the API in a standardized manner. Multiple instances of the server app are maintained using the \texttt{gunicorn}\footnote{\url{https://gunicorn.org/}} process manager.
The API supports a range of MongoDB Query Language (MQL)-based queries such as cone and general searches, map-reduce, and aggregation pipelines.

A dedicated \texttt{Kafka} consumer listens to the ZTF alert stream at IPAC and saves it to the database. It has the ability to filter and annotate the alerts prior to database ingestion (by e.g. evaluating ML models).

We choose to use \texttt{traefik}\footnote{\url{https://traefik.io}} as the reverse proxy/load balancer for its simplicity, performance, and encryption (TLS) support out-of-the-box.

\texttt{Kowalski} is containerized using the \texttt{Docker}\footnote{\url{https://docker.com}} software allowing for simple and efficient deployment in the cloud and/or on-premises.

\begin{figure*}
  \centering
  \subfigure[Search interface page]{\includegraphics[width=0.47\textwidth]{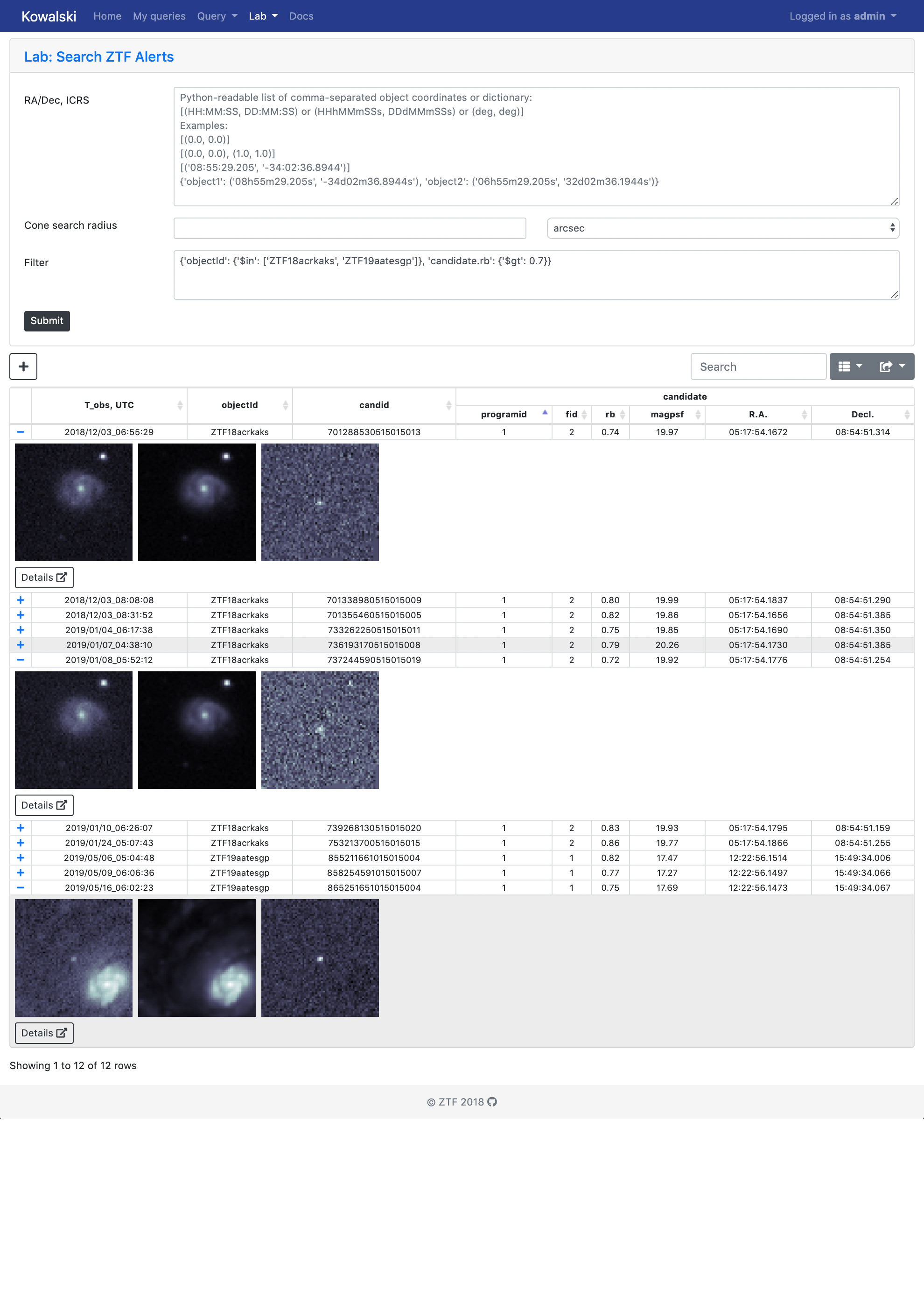}}\quad
  \subfigure[Individual alert page]{\includegraphics[width=0.47\textwidth]{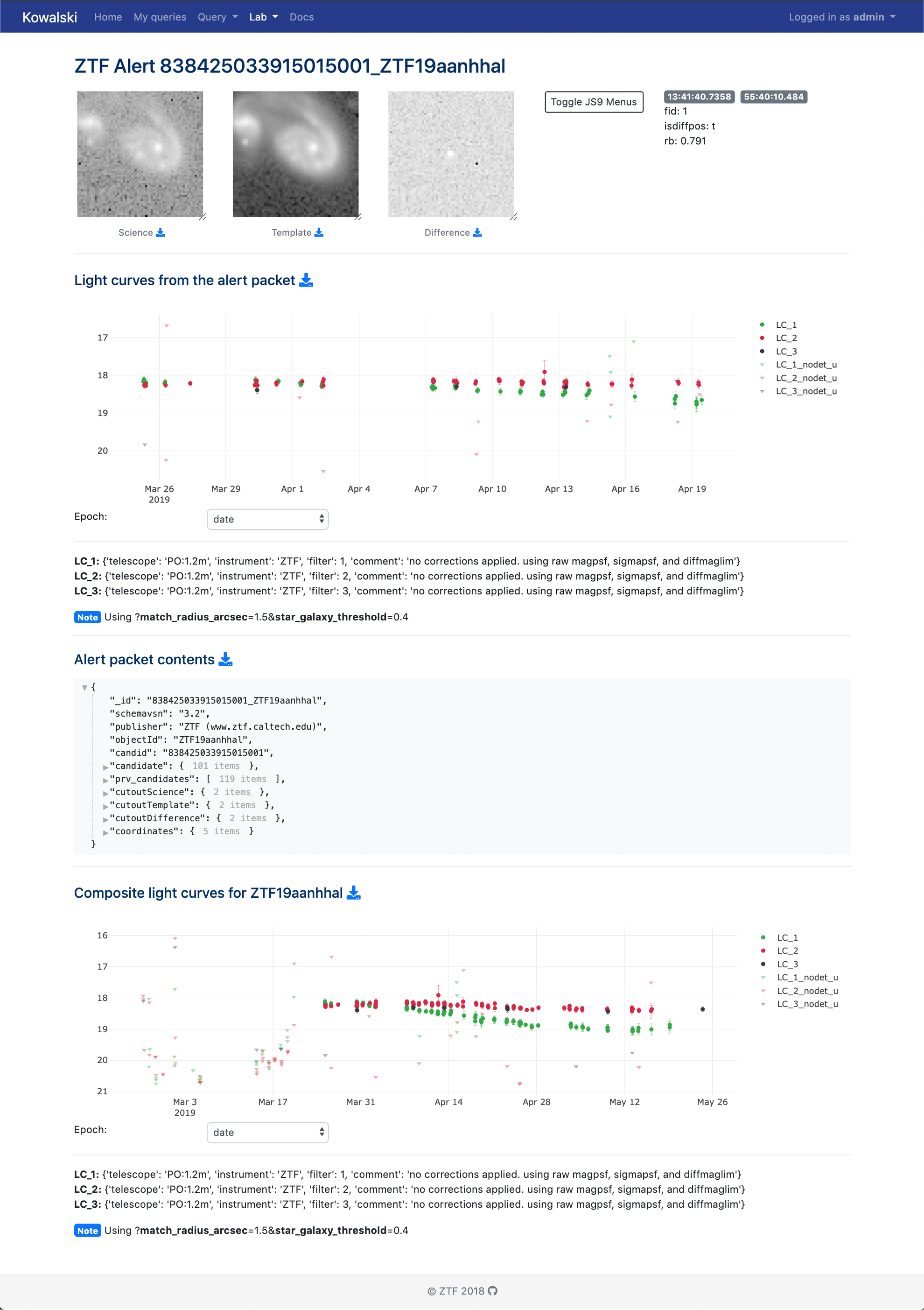}}
  \caption{\texttt{Kowalski}'s ZTF Alert Lab GUI.}
    \label{fig:Kowalski.alert_lab}
\end{figure*}

To simplify access to the ZTF alert data, \texttt{Kowalski} has a web-based GUI called the ZTF Alert Lab (ZAL), where users can efficiently search and preview alert contents (see Fig. \ref{fig:Kowalski.alert_lab}a). The ZAL also provides detailed views of individual alerts, interactively displaying the image cutouts (with \texttt{JS9}\footnote{\url{https://js9.si.edu/}}), alert contents, and the light curves (see Fig. \ref{fig:Kowalski.alert_lab}b). The latter may be corrected for the flux present in the reference (template) images.\footnote{The \textit{candidate.magpsf} field present in the alert packets reports the flux in the difference image, and is positive by construction. Alerts however may be from positive or negative subtractions (as identified by the \textit{candidate.isdiffpos} field), and for variable objects the flux in the reference image needs to be included.}

Additionally, the ZAL is able to construct compound object light curves since, due to packet size considerations, the individual alerts only contain a rolling 30-day window with historical data points.

As of June 2019, an instance of \texttt{Kowalski} deployed on-premises at Caltech stores over $30$ TB of various catalogs and databases including $125M+$ alerts and $2.5B+$ light curves. It processes millions of requests daily from $40+$ users, both programmatic services (e.g. ZTF's transient, variable, and Solar system marshals) and astronomers.

\subsubsection{Data labelling}

\begin{figure}
  \centering
  \includegraphics[width=0.47\textwidth]{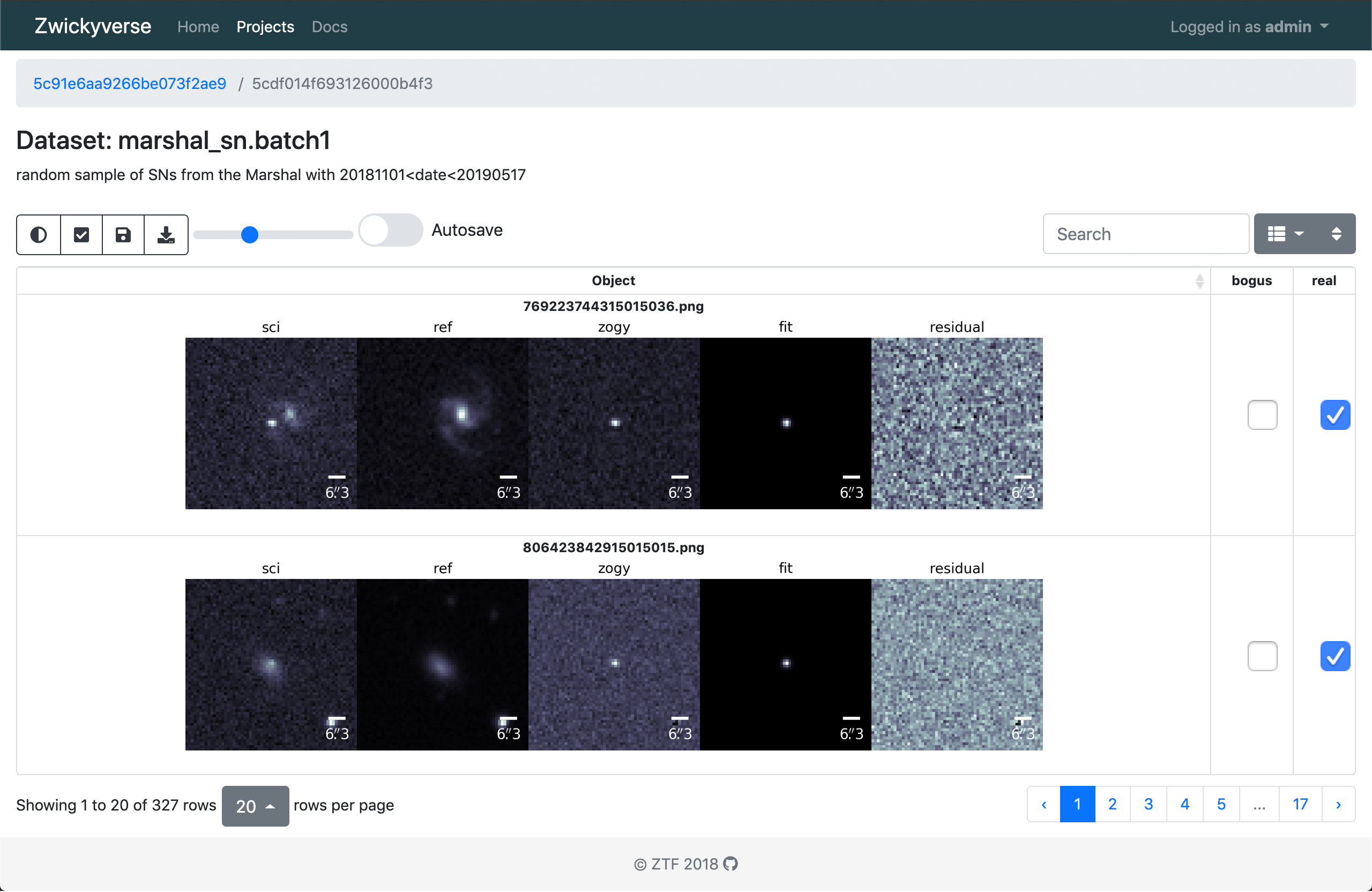}
    \caption{\texttt{Zwickyverse}'s GUI with sample image quintets (science, reference, difference, 2D-Moffat fit of difference, and 2D-Moffat fit minus difference residual)}
    \label{fig:Zwickyverse}
\end{figure}

For data labeling, we used a simple web-based open-source tool called \texttt{Zwickyverse}\footnote{\url{https://github.com/dmitryduev/zwickyverse}} that provides both an efficient API and GUI. The tool is easy to deploy thanks to containerization using \texttt{Docker} software and it allows quick integration of newly-labelled data sets into the model training workflow. All data labelling for this work was done using \texttt{Zwickyverse}.

To simplify labeling the image cutout triplets (science, reference, difference), we fitted the PSFs in the difference images\footnote{``D image'' in the ZOGY notation.} by 2D-Moffat functions and plotted those together with residual images (difference minus fit, see Fig. \ref{fig:Zwickyverse}). This proved to be helpful in identifying certain bogus detections. Additionally, contentious examples were individually inspected using the ZAL.

\subsubsection{Data diversity}

\begin{figure*}
  \centering
  \subfigure[Real examples: a supernova (top) and a variable star (bottom)]{\includegraphics[width=0.42\textwidth]{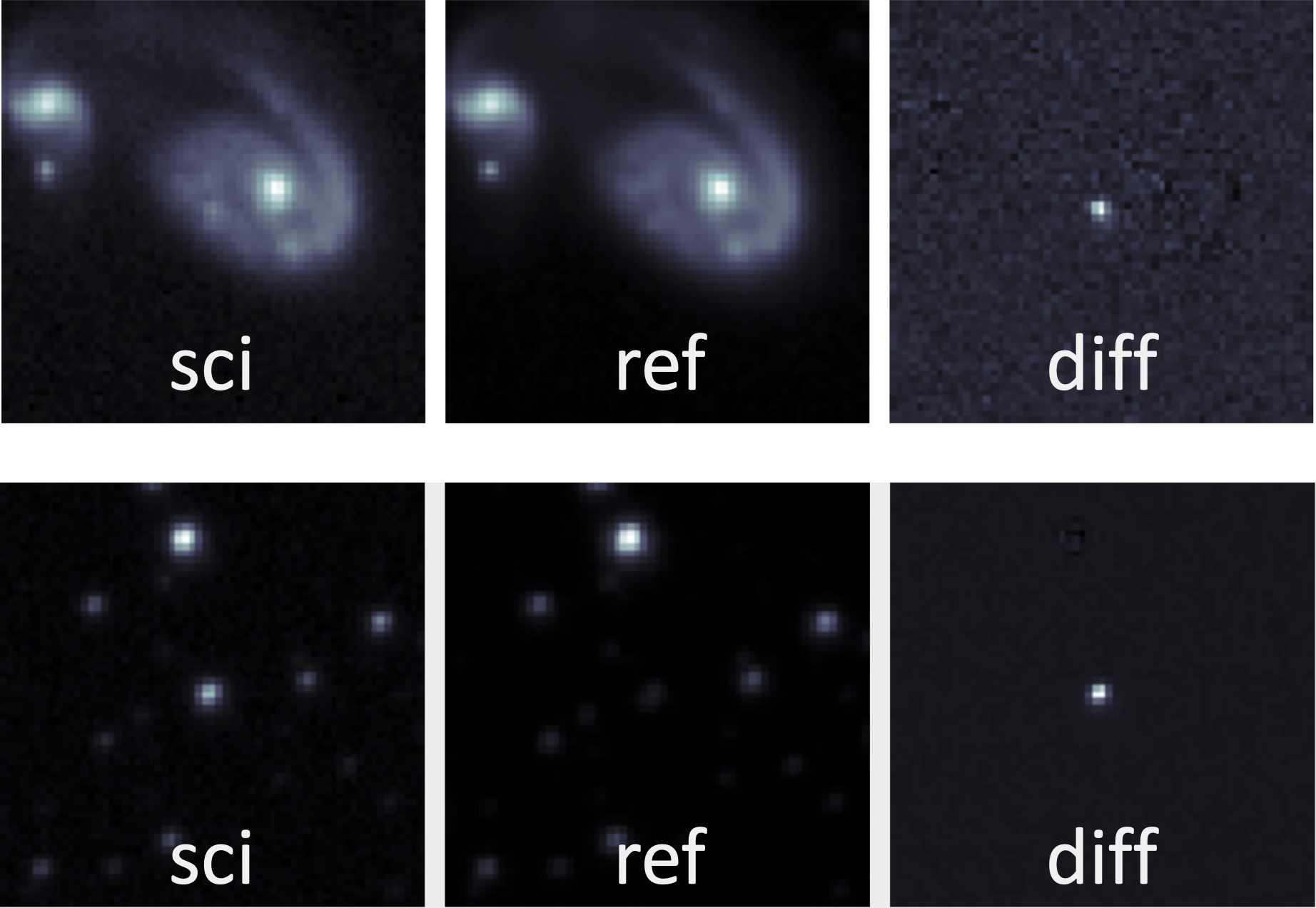}}\quad\quad
  \subfigure[Bogus examples: a badly-subtracted star (top) and an artifact caused by a masked bright object]{\includegraphics[width=0.42\textwidth]{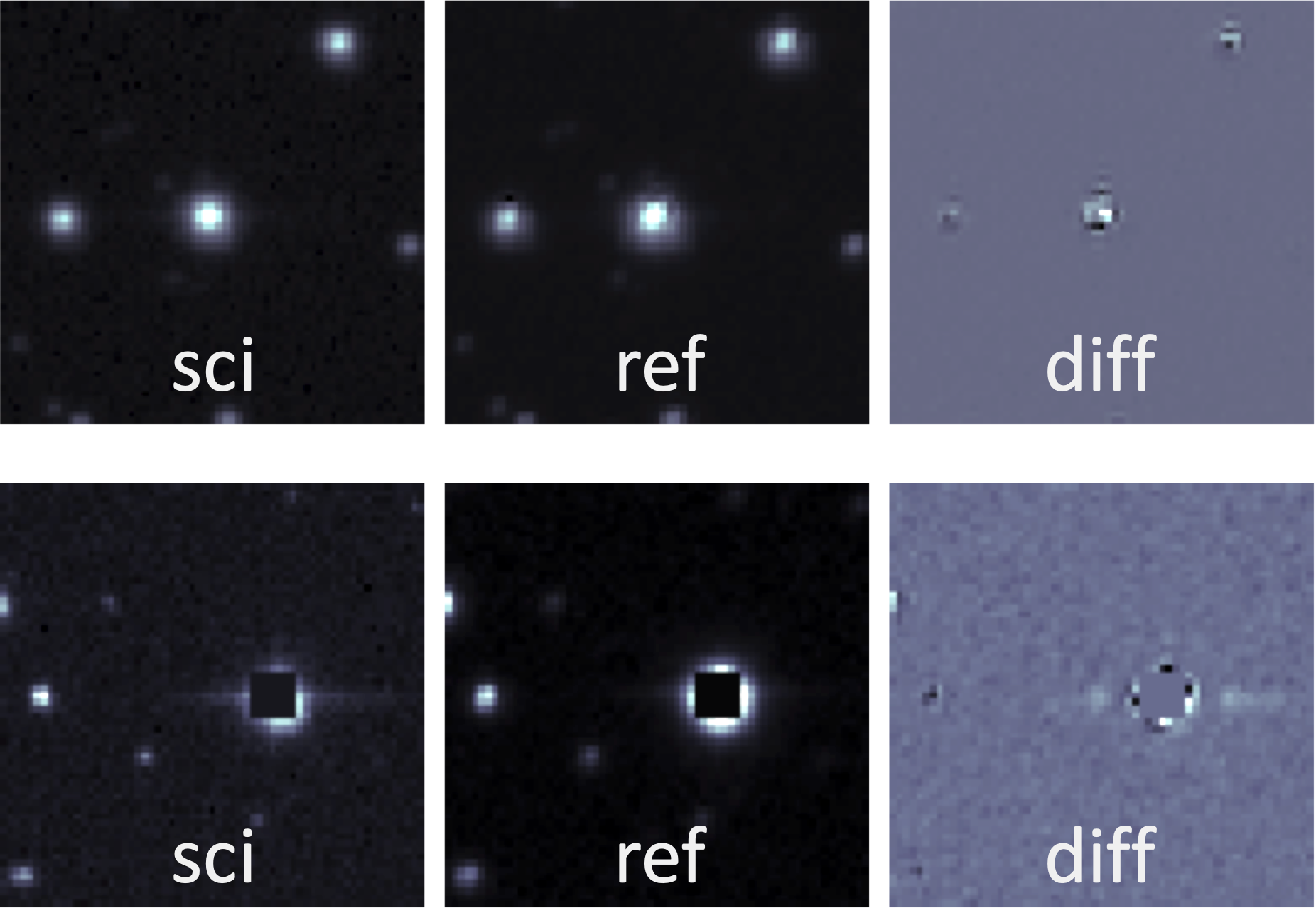}}
  \caption{Examples of 63x63 pixel cutout image triplets (science, reference, difference). ZTF plate scale is $1\arcsec$ per pixel.}
    \label{fig:real-bogus-triplets}
\end{figure*}

\begin{figure*}
  \centering
  \subfigure[Date]{\includegraphics[width=0.45\textwidth]{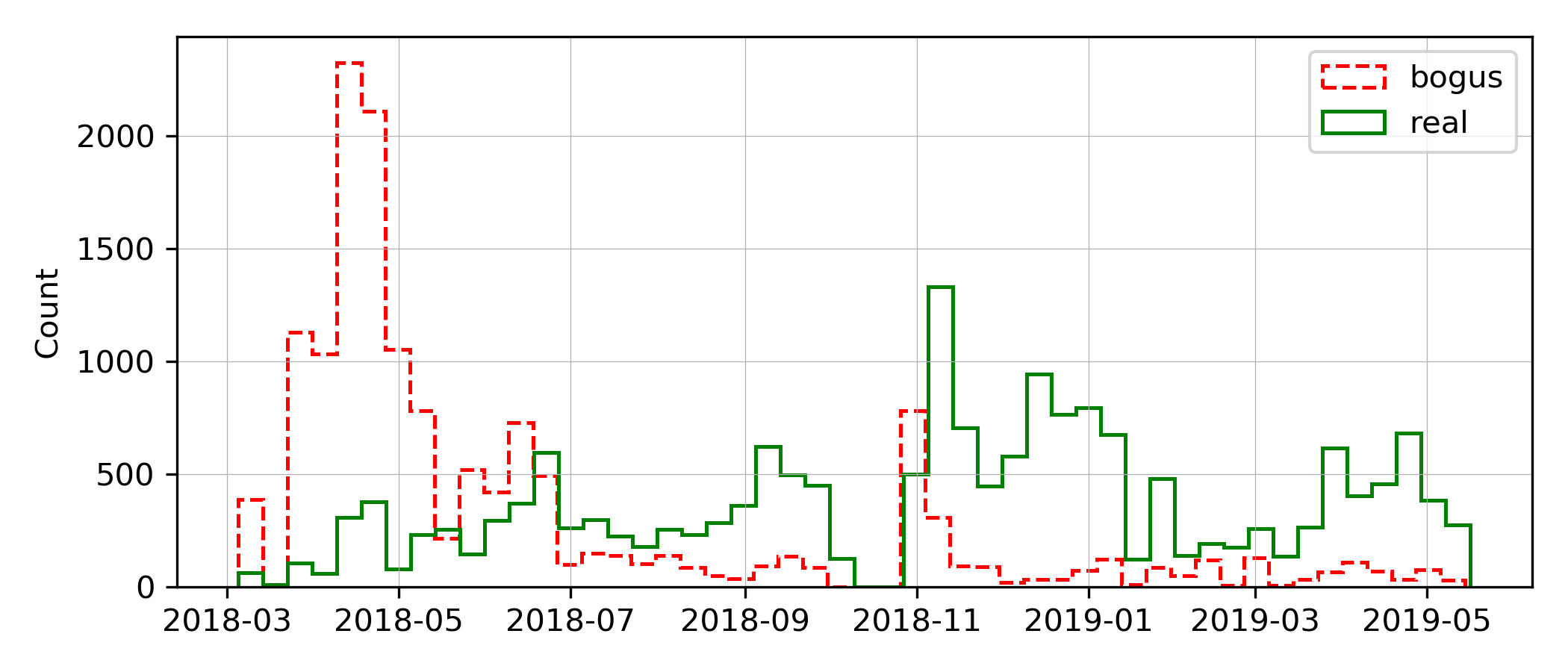}}\quad
  \subfigure[FWHM]{\includegraphics[width=0.45\textwidth]{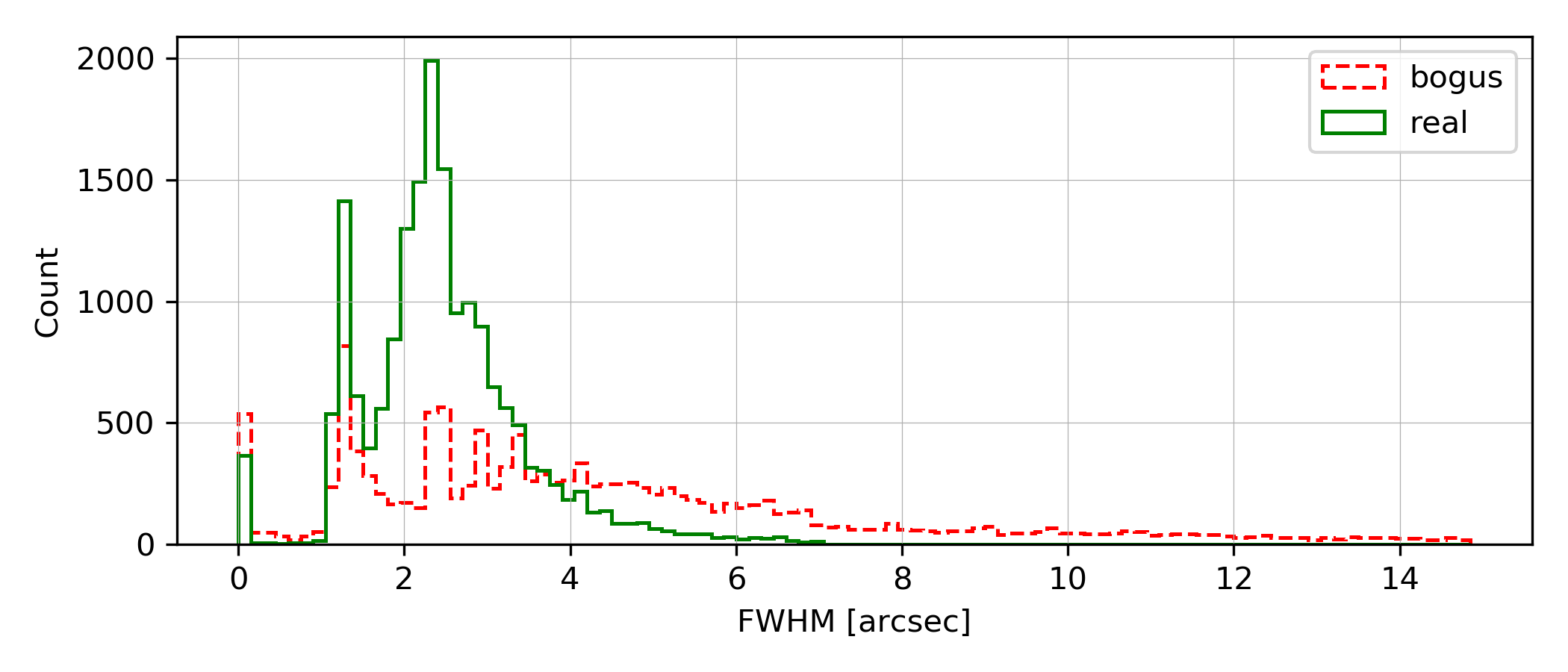}}\quad
  \subfigure[PSF magnitude]{\includegraphics[width=0.45\textwidth]{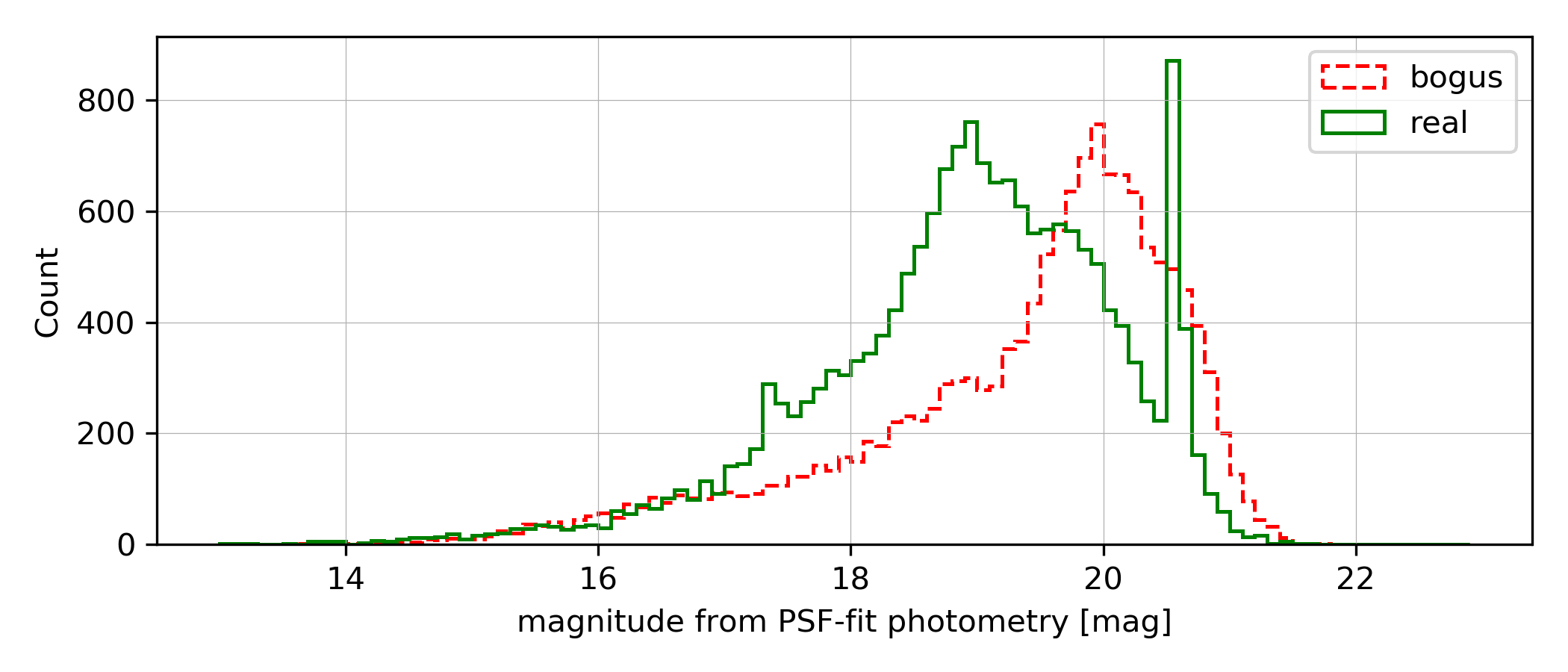}}\quad
  \subfigure[SNR]{\includegraphics[width=0.45\textwidth]{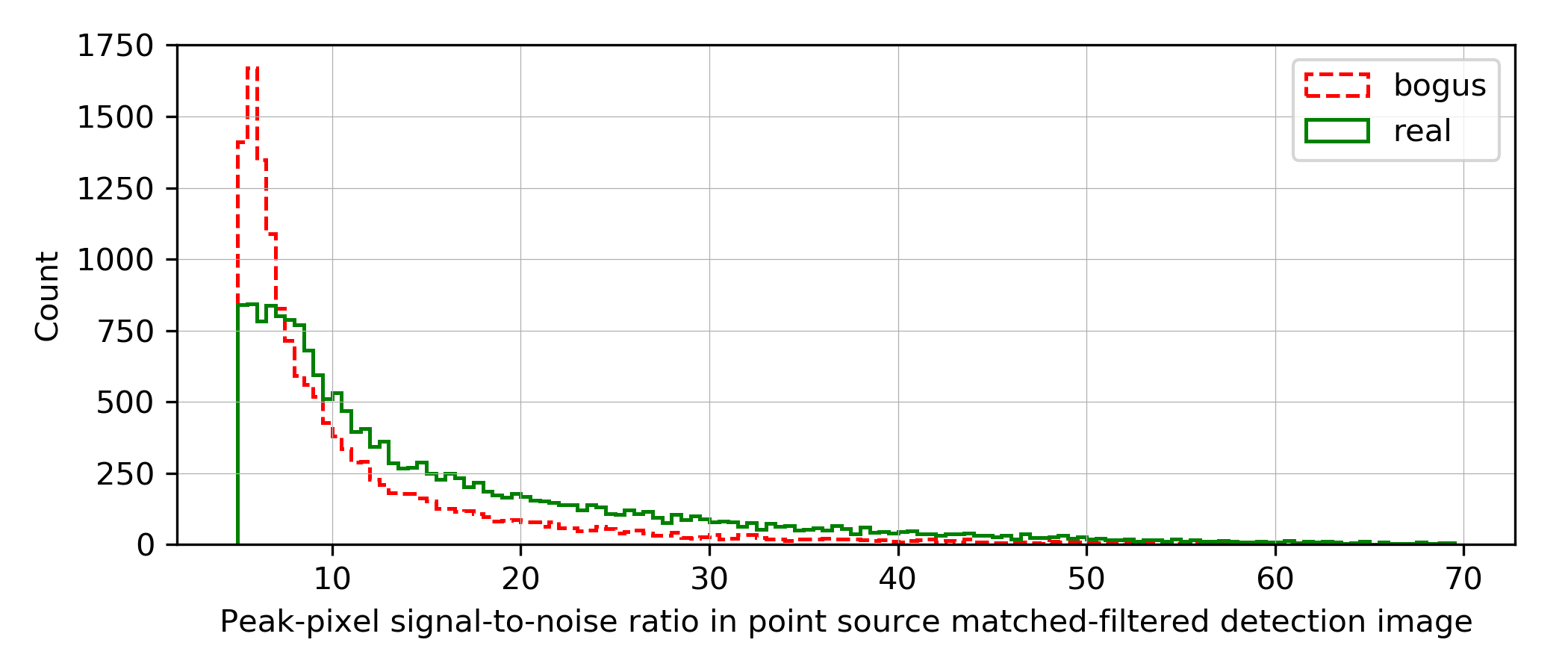}}
\caption{Training data breakdown by date, full width at half maximum (FWHM), PSF magnitude in the difference image, and peak-pixel signal-to-noise ratio in point source matched-filtered detection image. Most of the sharp peaks on the histograms are due to selection effects. The peak at zero on the FWHM histogram is due to a data processing artifact, which was fixed in May, 2018.}
\label{fig:diversity1}
\end{figure*}

\begin{figure*}
  \centering
  \subfigure[Filters]{\includegraphics[width=0.3\textwidth]{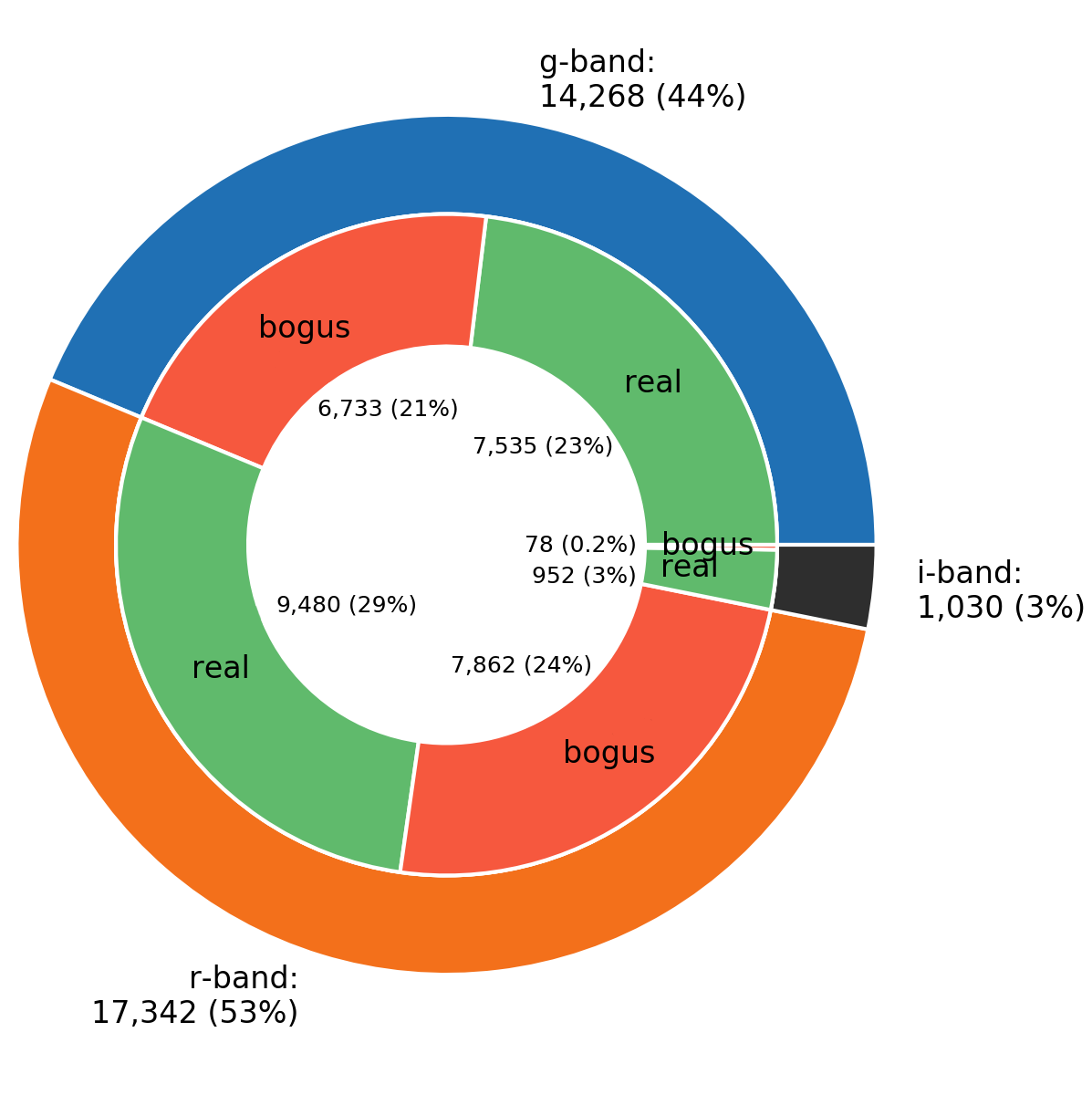}}\quad
  \subfigure[Galactic plane]{\includegraphics[width=0.3\textwidth]{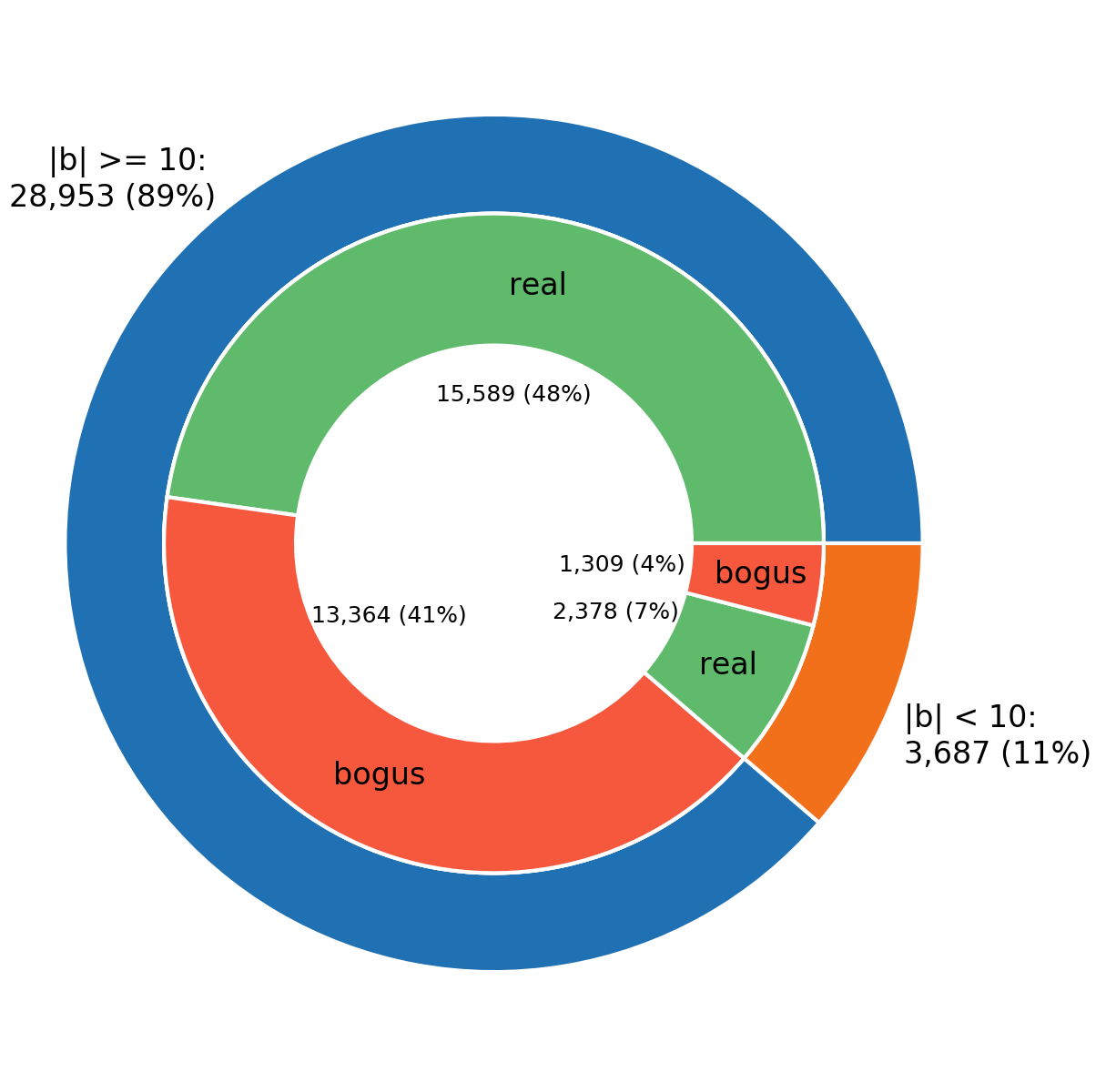}}\quad
  \subfigure[Positive/negative subtraction]{\includegraphics[width=0.3\textwidth]{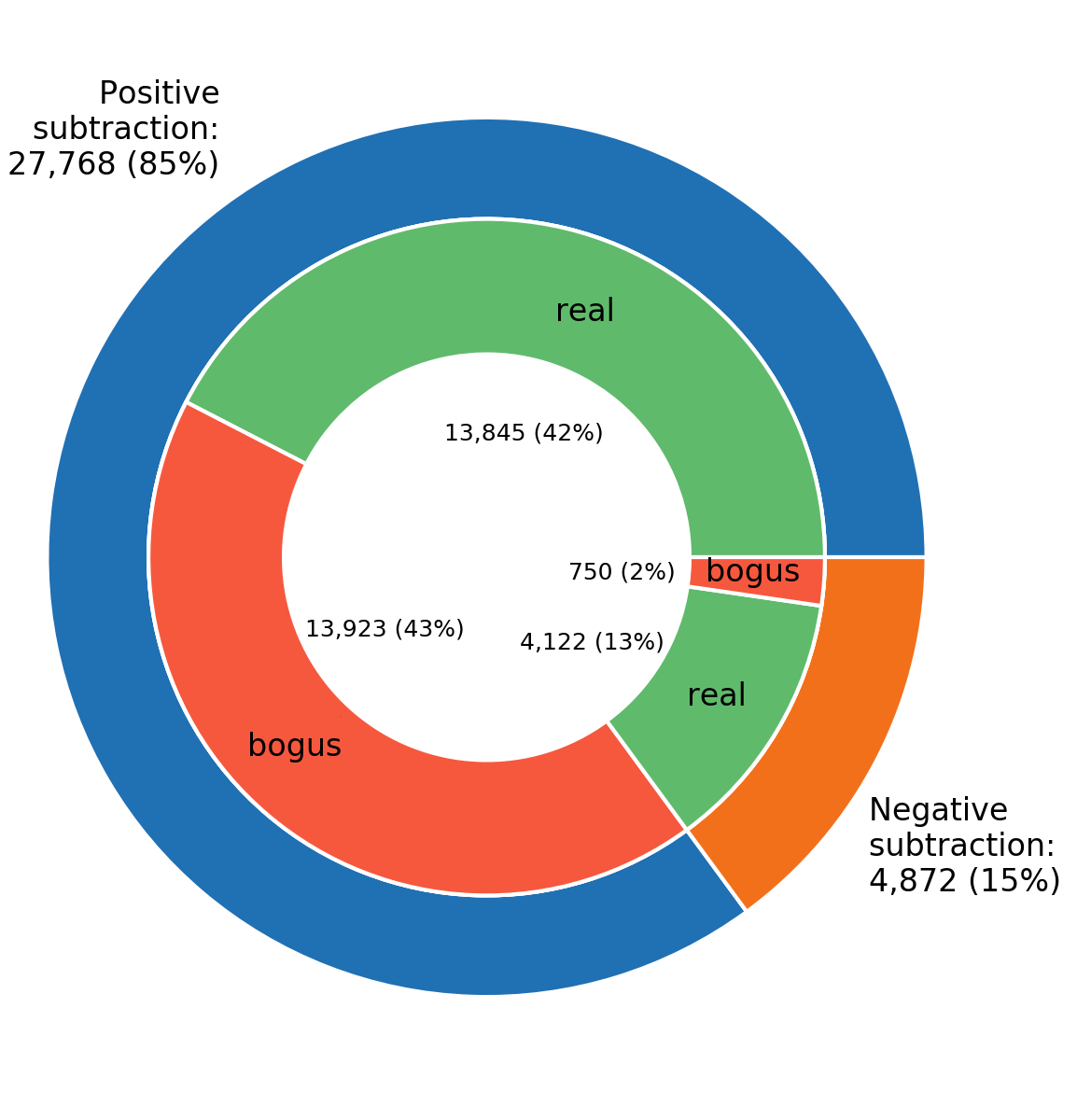}}
\caption{Training data breakdown by filters, position in/out of the Galactic plane, and positive/negative subtraction. Percentages of the total are given. As of June 2019, the total number of training examples is 32,640.}
\label{fig:diversity2}
\end{figure*}

We strove to build a data set that adequately samples the ZTF alert parameter space. For that, we collected over thirty  thousand training examples with the real-to-bogus data ratio of about $55\%/45\%$ (see Fig. \ref{fig:real-bogus-triplets}). Fig. \ref{fig:diversity1} shows the histograms of example counts as functions of the date and several candidate source characteristics extracted from the difference image: full width at half maximum (FWHM), PSF magnitude, and signal-to-noise ratio (SNR). Fig. \ref{fig:diversity2} shows the training set breakdown by filters, position in/out of the Galactic plane, and positive vs. negative subtractions. There are certain imbalances in the data set and we are planning to mitigate them in the future.

\subsection{braai architecture and training}\label{sec:braai_architecture}

\begin{figure*}
  \centering
  \includegraphics[width=0.97\textwidth]{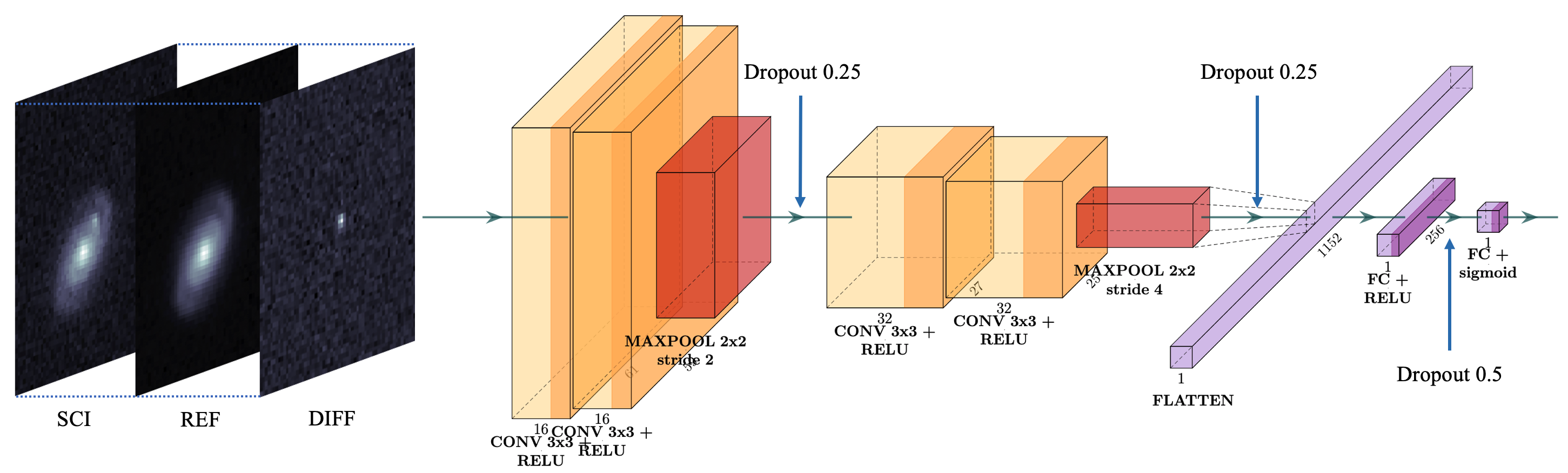}
    \caption{Architecture of the custom VGG6 model. The $L^2$-normalized epochal science, reference, and difference cutout images are stacked to form 63x63x3 triplets that are input into the model. ReLU activation functions are used for all five hidden trainable layers; a sigmoid activation function is used for the output layer that produces a score from 0.0 to 1.0. Dropout is used for regularization. See Section \ref{sec:braai_architecture} for the details.}
    \label{fig:vgg}
\end{figure*}

We use a simple custom VGG\footnote{This architecture was first proposed by the Visual Geometry Group of the Department of Engineering Science, University of Oxford, UK}-like sequential model (``VGG6'') \citep{2014arXiv1409.1556S} (see Fig. \ref{fig:vgg} for details). The model has six layers with trainable parameters: four convolutional and two fully-connected. The first two convolutional layers use 16 3x3 pixel filters each while in the second pair, 32 3x3 pixel filters are used. To prevent over-fitting, a dropout rate of 0.25 is applied after each max-pooling layer and a dropout rate of 0.5 is applied after the second fully-connected layer. ReLU activation functions\footnote{Rectified Linear Unit --  a function defined as the positive part of its argument} are used for all five hidden trainable layers; a sigmoid activation function is used for the output layer.

We also implemented more complicated architectures such as 18- and 50-layer deep models based on residual connections (``ResNet18'' and ``ResNet50''), but observed no performance gain and therefore only use the VGG6 model.

The cutout images that are generated by the ZSDS are centered on the event candidate and are of size 63x63 pixels (or smaller, if the event is detected near the CCD edge) at a plate scale of $1\arcsec$ per pixel. We perform independent $L^2$-normalization of the epochal science, reference, and difference cutouts and stack them to form 63x63x3 triplets that are input into the model. Smaller examples are accordingly padded using a constant pixel value of $10^{-9}$.

\texttt{Braai} is implemented using \texttt{TensorFlow} software and its high-level \texttt{Keras} API \citep{tensorflow2015-whitepaper, chollet2015keras}. We used the binary cross-entropy loss function, the Adam optimizer \citep{2014arXiv1412.6980K}, a batch size of 64, and a 81\%/9\%/10\% training/validation/test data split. The training image data were weighted per class to mitigate the slight real vs. bogus imbalance in the data sets. The images may be flipped horizontally and/or vertically at random. No random rotations and translations were added. 

\begin{figure}
  \centering
  \includegraphics[width=0.45\textwidth]{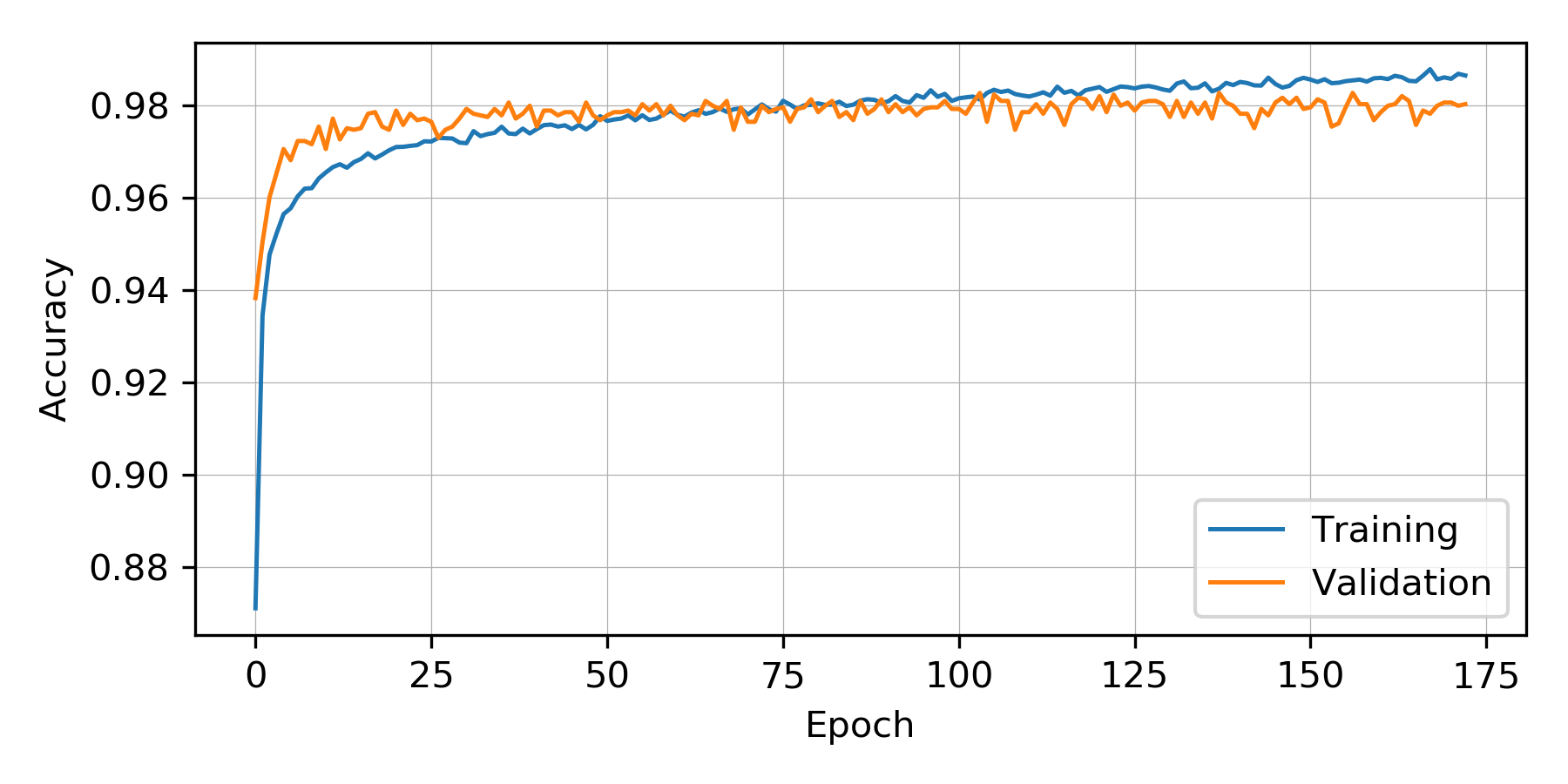}
    \caption{Training (in blue) and validation (in orange) accuracy of \texttt{braai} version $d6\_m7$ that is deployed in production as of June 2019.}
    \label{fig:accuracy}
\end{figure}

We used the early stopping technique to finish training if no improvement in validation accuracy was observed over many epochs. As a result, the model is typically trained for 150-200 epochs. For training, we used an on-premises Nvidia Tesla P100 12G GPU. Training for 200 epochs on $\sim30k$ images takes about 20 minutes for the VGG6 architecture.

Figure \ref{fig:accuracy} shows training (in blue) and validation (in orange) accuracy for the model version $d6\_m7$ that is deployed in production as of June 2019. Both the training and validation accuracy is about 98\%.

\begin{figure}
  \centering
  \includegraphics[width=0.47\textwidth]{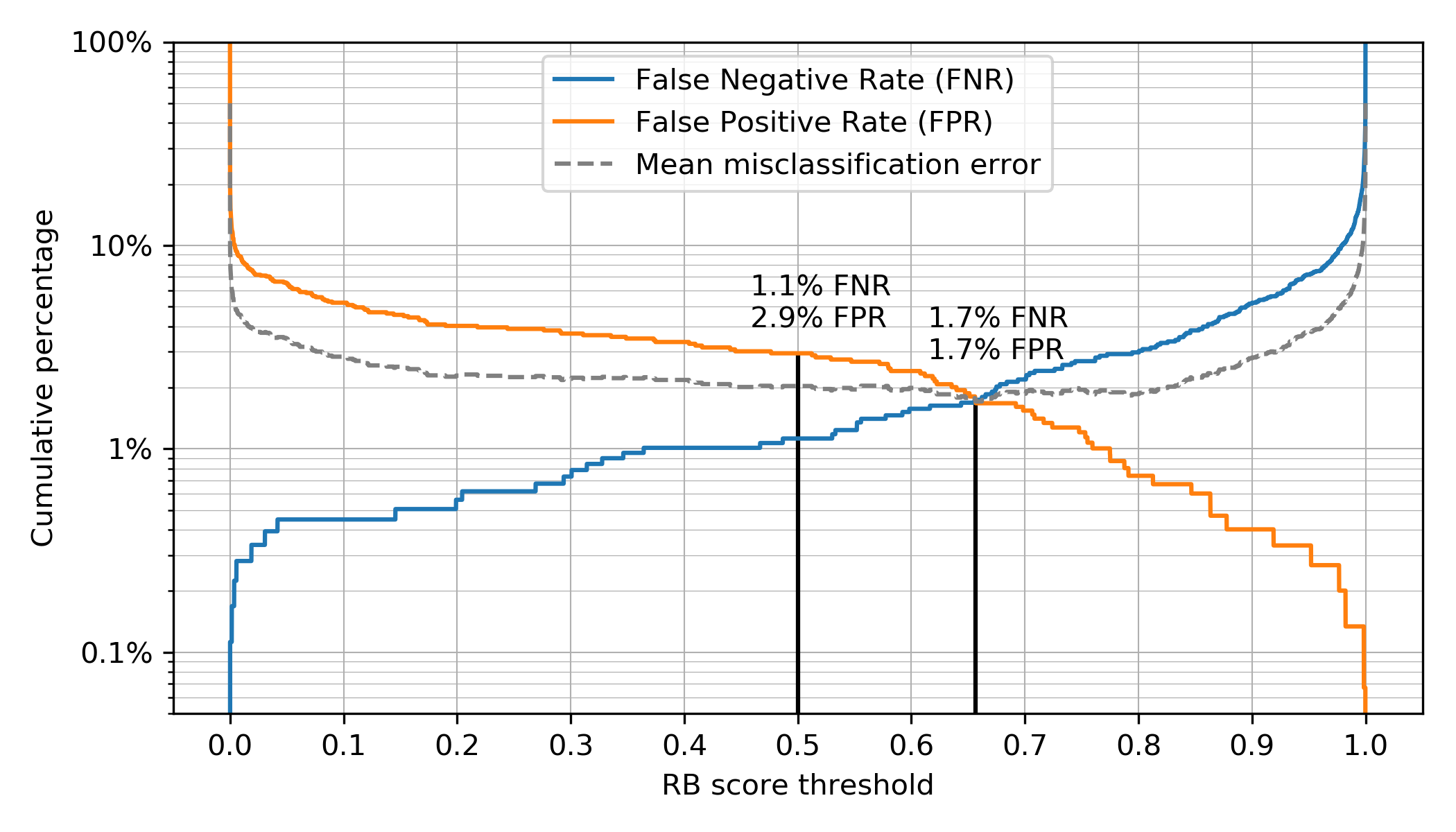}
    \caption{FNR and FPR as functions of the RB score threshold. At a score threshold of 0.5, \texttt{braai} yields $1.1\%$ FNR and $2.9\%$ FPR. At a score threshold of 0.65, \texttt{braai} yields a value of $1.7\%$ for both FNR and FPR. \texttt{Braai} version $d6\_m7$ (deployed in production as of June 2019) was evaluated on 3,271 test examples from the data set.}
    \label{fig:performance}
\end{figure}

\begin{figure}
  \centering
  \includegraphics[width=0.42\textwidth]{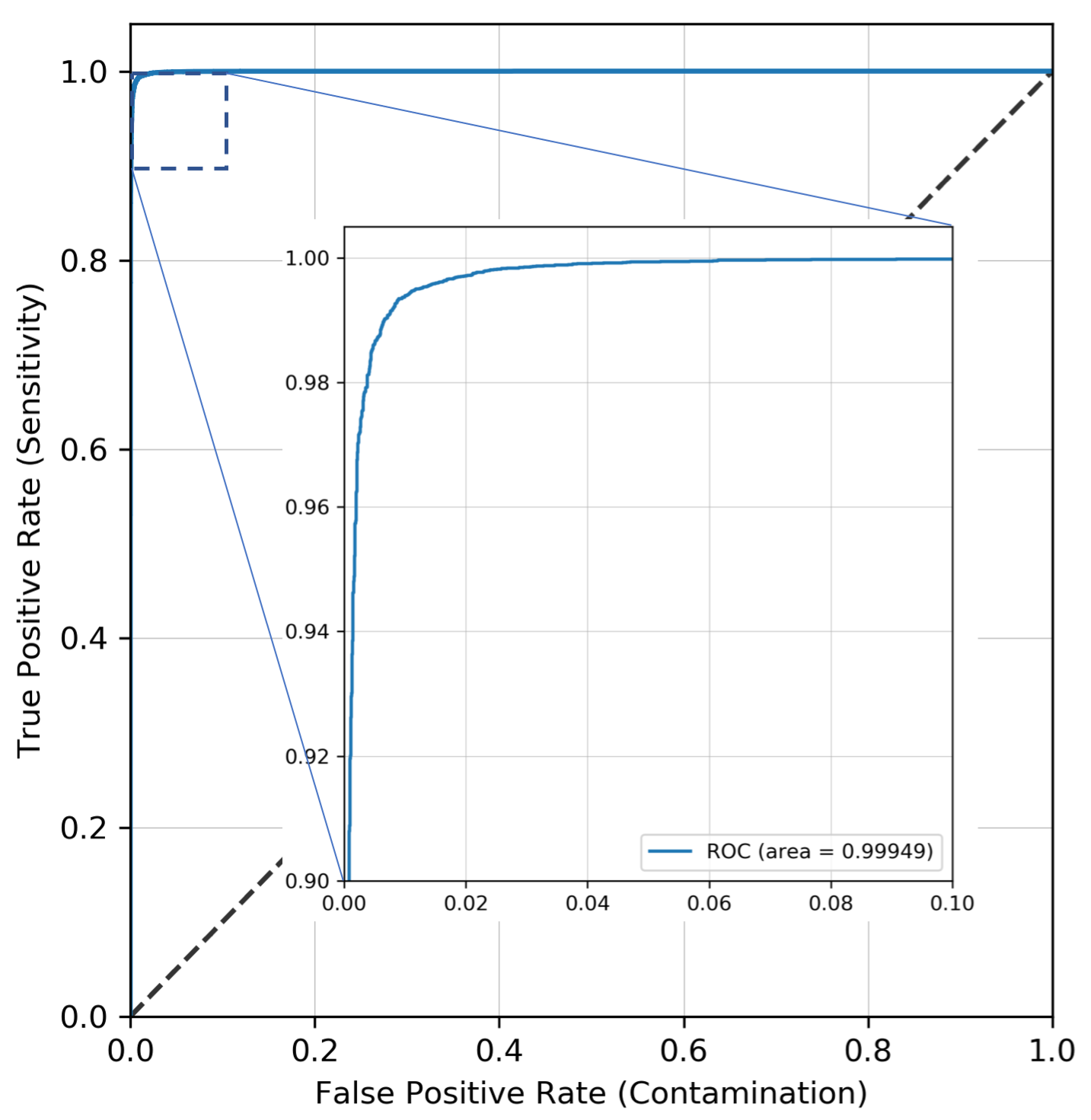}
    \caption{ROC curve of \texttt{braai} version $d6\_m7$ that is deployed in production as of June 2019.}
    \label{fig:roc}
\end{figure}

\begin{figure}
  \centering
  \subfigure[Without normalization]{\includegraphics[width=0.2\textwidth]{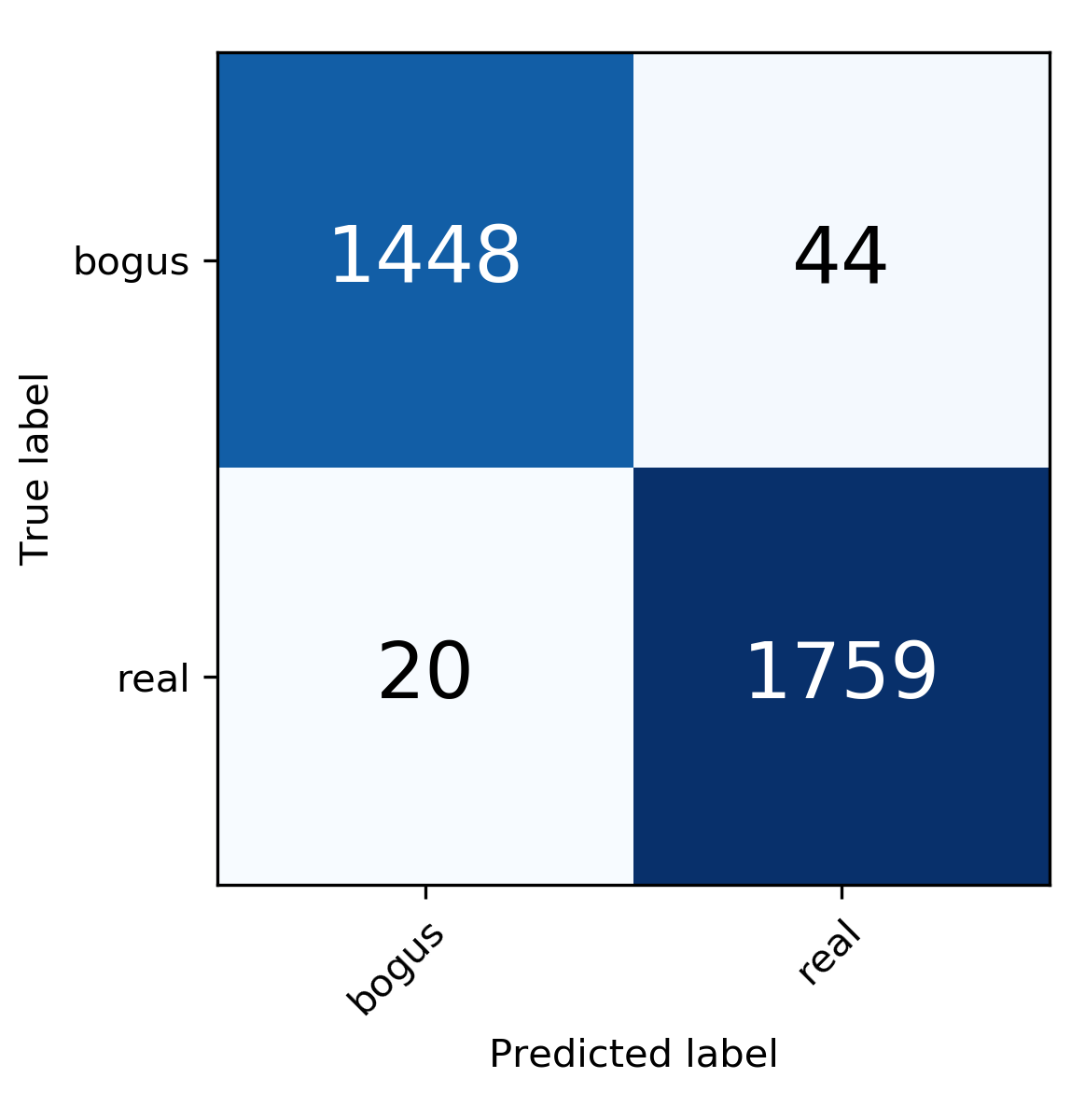}}\quad
  \subfigure[Normalized]{\includegraphics[width=0.2\textwidth]{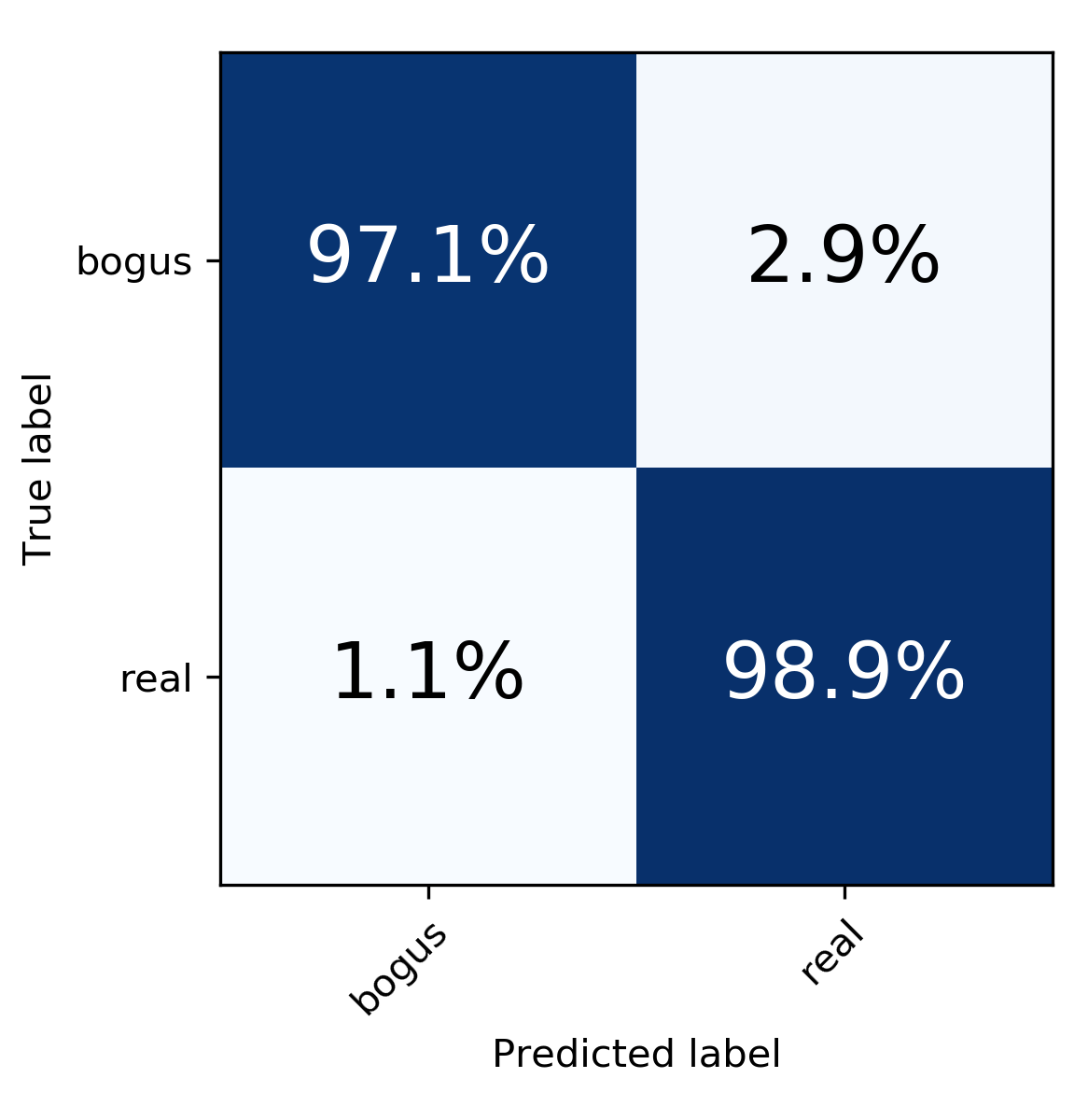}}
\caption{Confusion matrices for an RB score threshold of 0.5 for \texttt{braai} version $d6\_m7$ that is deployed in production as of June 2019.}
\label{fig:cm}
\end{figure}

The test performance of the resulting classifier quantified by the false negative rate (FNR) and false positive rate (FPR) as functions of the score threshold is shown in Fig. \ref{fig:performance}. Fig. \ref{fig:roc} displays the receiver operating characteristic (ROC) curve. At a score threshold of 0.5, \texttt{braai} yields $1.1\%$ FNR on the test set (which contained 3,271 examples) while keeping the FPR below $3\%$, as demonstrated in the confusion matrices (Fig. \ref{fig:cm}). At a score threshold of 0.65, \texttt{braai} yields a value of $1.7\%$ for both FNR and FPR.

\section{Classifier performance}

\subsection{Zooniverse test set}\label{sec:test_sets}

Zwicky's Quirky Transients is a Zooniverse project for volunteers to publicly label a set of ZTF candidates as real or bogus (with a skip option). We show only three images per candidate viz. Sci, Ref, and Diff. from which the volunteers are expected to tell apart the two types. One thing we wanted to see was the consensus especially where RF RB scores were ambiguous. Up to ten volunteers could classify the same object. The volunteers were trained through tutorials, a field guide, and by feedback from researchers in interactive Q\&A threads. The RF RB score ranges were hand-crafted for them to show examples that were definitely bogus, definitely real, and ambiguous (majority).

Seven campaigns were run between January and May of 2019 involving $\sim 13000$ objects. 
The first campaign was the largest with 6600 triplets with the following RF $rb$ distribution: $10\%\: rb < .3$, $10\%\:  rb > .7$, $80\%\: 0.3 <= rb <= 0.65$. The second and third campaigns had similar $rb$ distributions, and $\sim 1000$ objects each. The last four campaigns also had $\sim 1000$ objects each, but all had $rb > 0.4$. For all the campaigns we had excluded objects within 8'' of known Solar System objects, and also those that have been found through subtracting the Sci image from the Ref image (i.e. with a fainter Sci image detection compared to the Ref image detection). At the end of the first campaign, we selected a data set consisting of triplets with at least six separate classifications.  For this set of 6436, we plotted the $rb$ score against the fraction of real classifications, and selected two areas from the plot -- (a) ``gold real'': those with $rb$ score $>= 0.57$ and classified by at least $70\%$ volunteers as real, and (b) ``gold bogus'': those with $rb$ score $ <= 0.45$ and  classified by fewer than $45\%$ volunteers as real. This resulted in the gold real set having 416 triplets and, the gold bogus having 1,196 triplets, for a total of 1612 triplets.  These data were then inspected using \texttt{Zwickyverse} and the ZAL and re-labeled if necessary. We found that the real event sample was almost uncontaminated (under 5\% FPR), however about 30\% of what was labeled as bogus (and also had low RF RB scores) turned out to be real (mainly, variable stars imaged under challenging conditions, something the volunteers could not have known because they had no access to the light curves).

\begin{figure}
  \centering
  \includegraphics[width=0.47\textwidth]{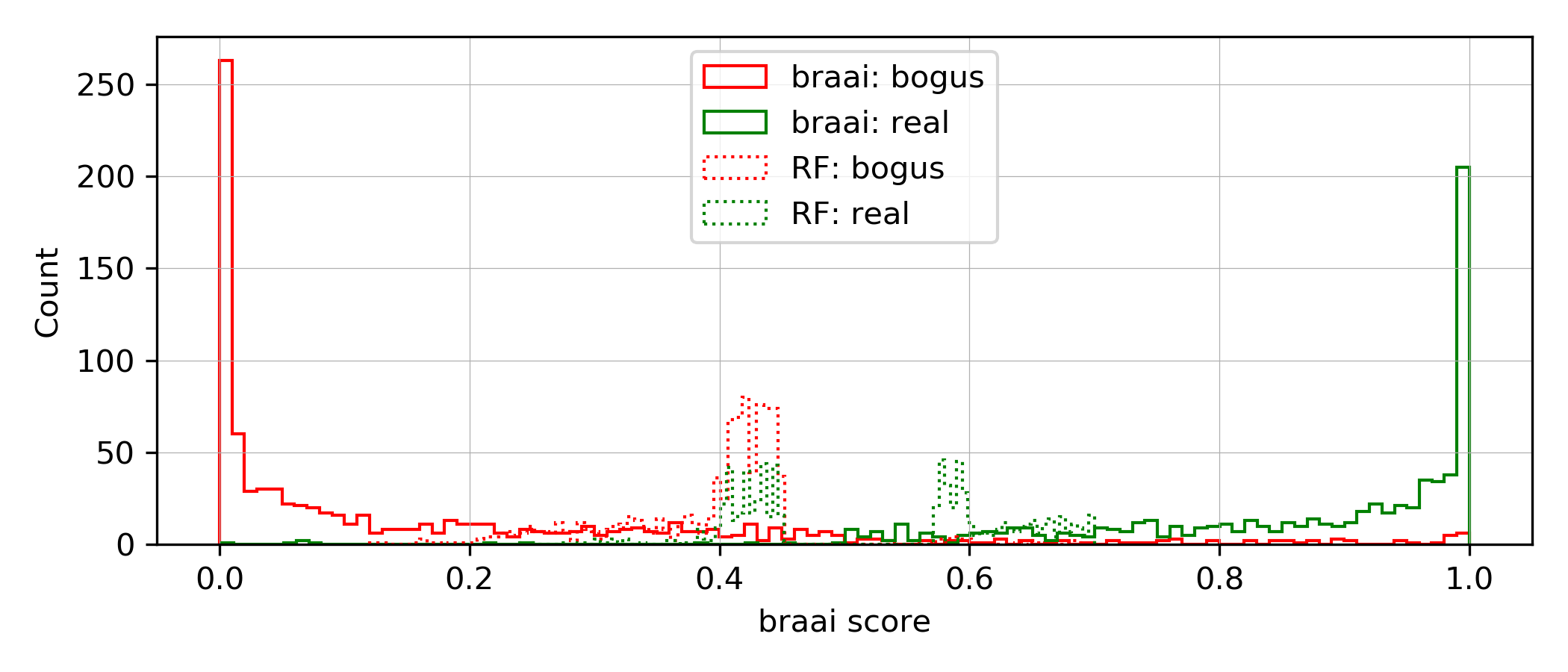}
    \caption{Histogram of \texttt{braai} and RF RB scores for 734 real and 878 bogus examples from the Zooniverse test set. \texttt{Braai} version $d6\_m7$ that is deployed in production as of June 2019. RF RB scores come from multiple versions of the classifier.}
    \label{fig:zoo}
\end{figure}

Fig. \ref{fig:zoo} shows the histogram of \texttt{braai} (version $d6\_m7$ deployed in production as of June 2019) and RF RB scores (that come from multiple versions of the classifier) for 734 real and 878 bogus examples from the Zooniverse test set. Evidently, \texttt{braai} yields much more reliable results than the RF classifier. The two peaks around RF RB of 0.5 are a selection effect caused by the input ranges of RB scores chosen for the Zooniverse campaigns.

\subsection{Real events}

\begin{figure}
  \centering
  \includegraphics[width=0.45\textwidth]{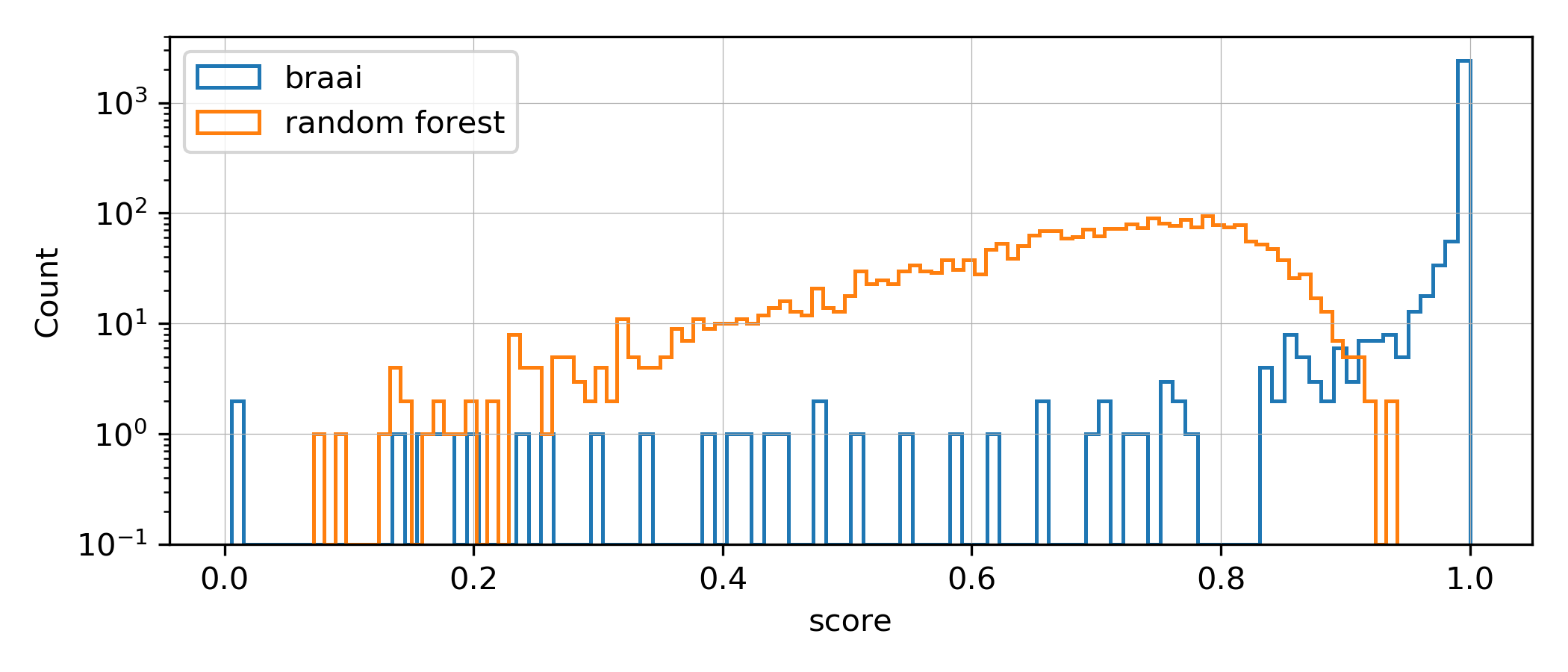}
    \caption{Histogram of \texttt{braai} and RF RB scores of 2,633 ZTF alerts from the night of May 14, 2019 that originated from 921 objects identified as real on the GROWTH marshal. \texttt{Braai} version $d6\_m7$. RF classifier version $t15\_f5\_c3$.}
    \label{fig:reals_20190514}
\end{figure}

To further test the performance, we evaluated \texttt{braai} on 2,633 ZTF alerts from the night of May 14, 2019 that originated from 921 objects vetted as real by humans on the GROWTH marshal after passing programmatic filters of different science groups (a mix of flux-transient and reoccurring flux-variable objects).\footnote{The GROWTH marshal users have additional information available to them such as spectroscopic follow-up and cross-matches to external surveys.} Again adopting a score threshold of 0.5, only 18 out of 2633 candidates are misclassified (0.7\% FNR) by \texttt{braai} compared to 282 (10.7\% FNR) misclassified by the RF classifier deployed at the time. The histogram of \texttt{braai} and RF RB scores is shown in Fig. \ref{fig:reals_20190514} in logarithmic scale, since the vast majority of these candidates are scored close to unity by \texttt{braai}.

\begin{figure}
  \centering
  \includegraphics[width=0.45\textwidth]{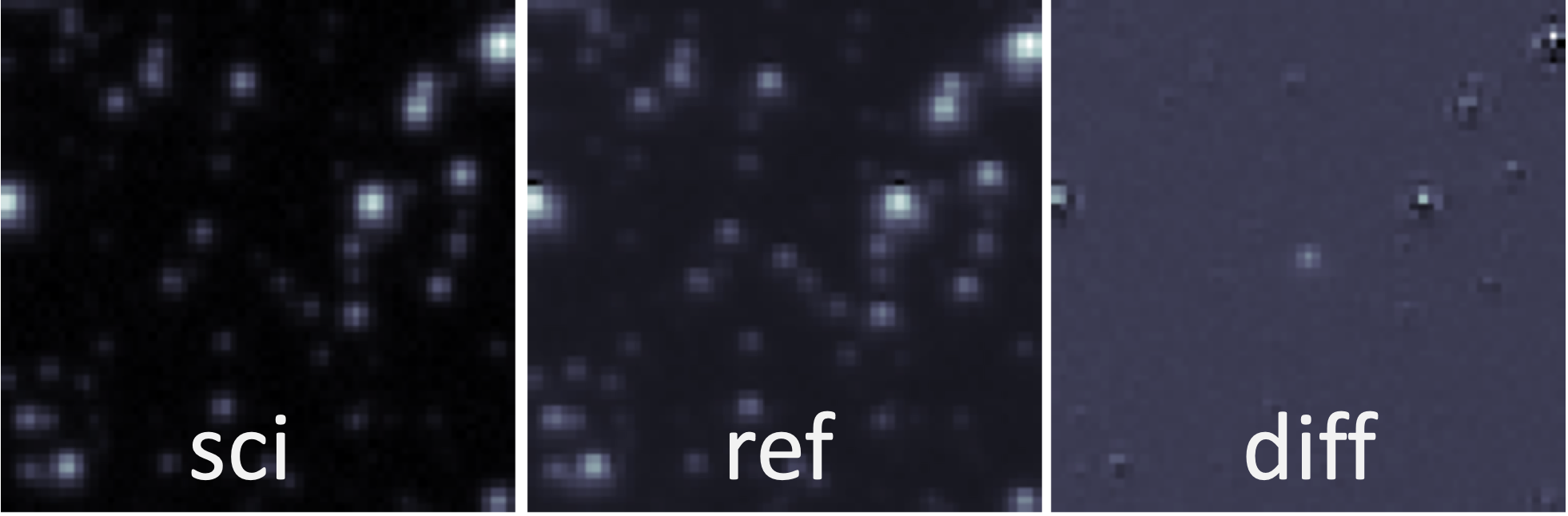}
    \caption{Example 63x63 pixel image triplet (science, reference, difference) from a known reoccurring flux-variable object located in a densely populated region of the sky. ZTF plate scale is $1\arcsec$ per pixel.}
    \label{fig:triplet_dense}
\end{figure}

\begin{figure}
  \centering
  \includegraphics[width=0.45\textwidth]{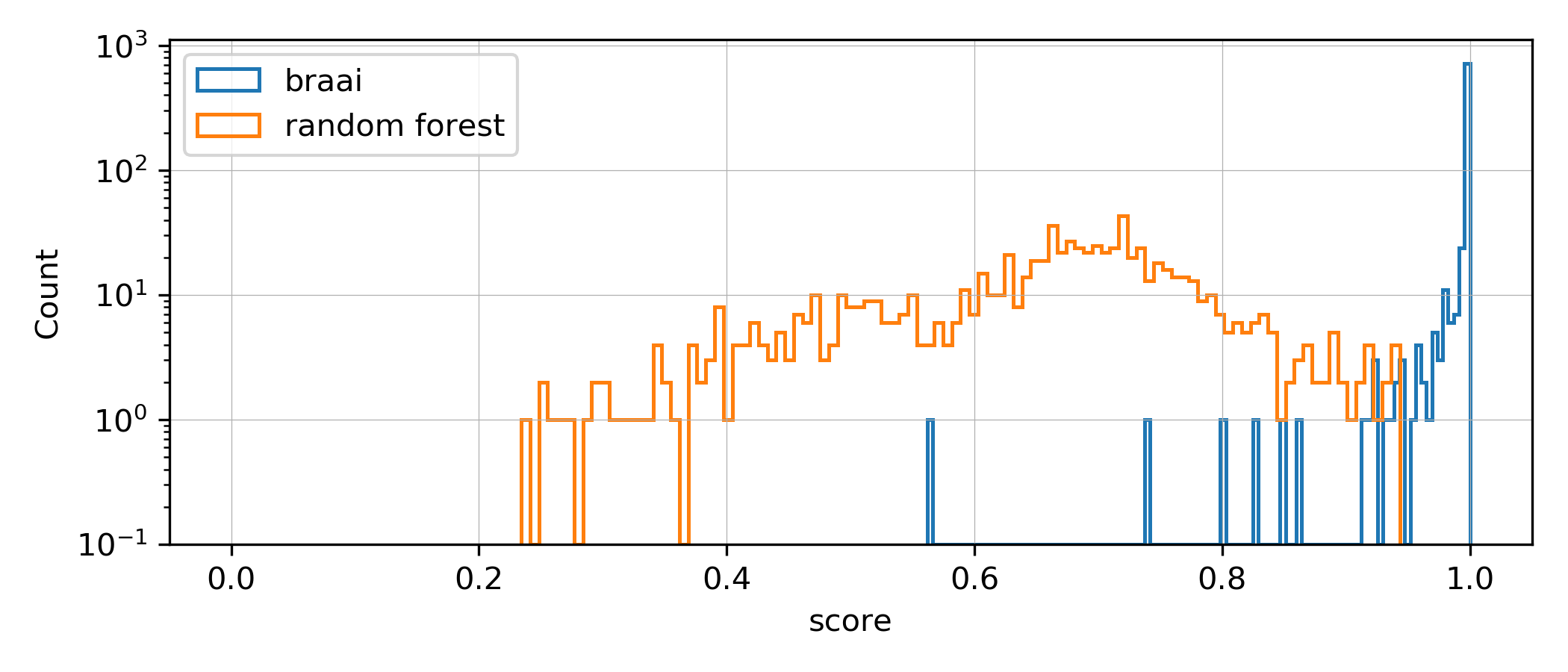}
    \caption{Histogram of \texttt{braai} and RF RB scores of 803 ZTF alerts that originated from objects located in densely populated regions of the sky. \texttt{Braai} version $d6\_m7$ deployed in production as of June 2019. RF RB scores come from multiple versions of the classifier.}
    \label{fig:dense}
\end{figure}

Next, we tested the \texttt{braai} performance on 803 alerts from known reoccurring flux-variable objects located in densely populated regions of the sky (see an example triplet on Fig. \ref{fig:triplet_dense}). The alerts were generated from observations covering a wide range of conditions over the course of ZTF's first year of operation. With a score threshold of 0.5, all candidates are classified correctly by \texttt{braai} compared to 113 misclassifications (14.1\% FNR) by the RF classifier. The histogram of \texttt{braai} and RF RB scores is shown in Fig. \ref{fig:dense} in logarithmic scale.

\begin{figure}
  \centering
  \includegraphics[width=0.45\textwidth]{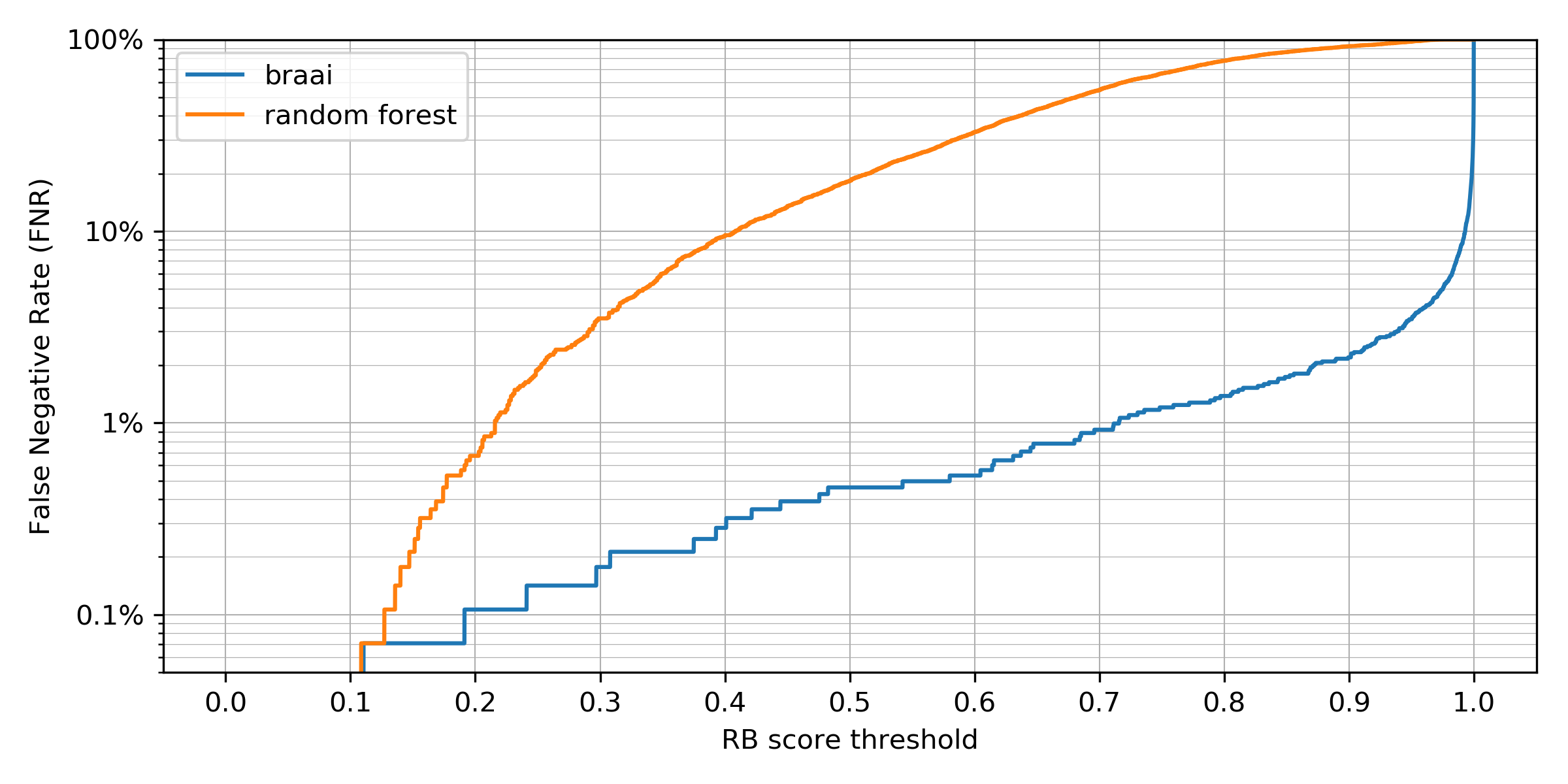}
    \caption{False negative rates of \texttt{braai} and RF RB classifier for 2,820 alerts from a set of 140 supernovae detected by ZTF in 2019. At a score threshold of 0.5, \texttt{braai} correctly identified 99.5\% of the SN vs. 80\% by RF RB. \texttt{Braai} version $d6\_m7$ deployed in production as of June 2019. RF classifier version $t15\_f5\_c3$ or more recent.}
    \label{fig:sn}
\end{figure}

Further, we evaluated \texttt{braai} on 2,820 alerts from a set of 140 recent (detected in 2019) supernovae (SNs). We selected those SNs such that no alerts originating from them were in the training set. As shown in Fig. \ref{fig:sn}, the FNR of \texttt{braai} stays below 1\% up to a score threshold of 0.7 essentially ensuring detection of (even very young) SNs while keeping the FPR below 2\% (see Fig. \ref{fig:performance}), significantly reducing the amount of time spent on SN candidate vetting.

\begin{figure*}
  \centering
  \subfigure[Histogram of Solar system object V-band magnitudes in  the set, as reported by the Minor Planet Center archive. The objects are identified automatically, resulting in a set that is not limited to high signal-to-noise levels that humans usually find easy to detect.]{\includegraphics[width=0.47\textwidth]{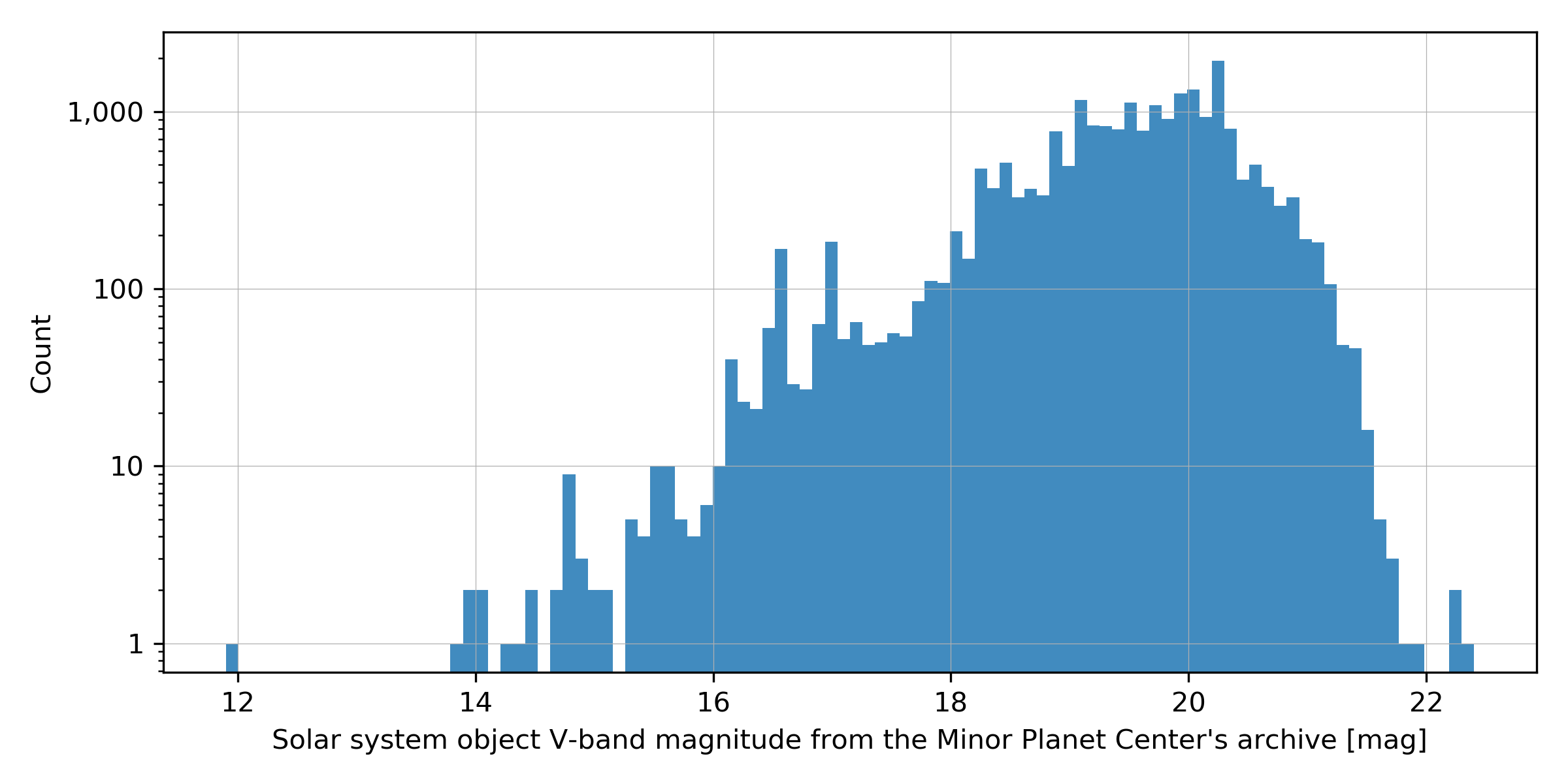}}\quad
  \subfigure[False negative rates of \texttt{braai} vs. RF RB classifier for asteroids with $Vmag < 18.5$ (2,534 alerts), $18.5 <= Vmag < 20.5$ (16,587 alerts), and $Vmag >= 20.5$ (2,520 alerts). At a score threshold of 0.5, \texttt{braai}'s FNR stays below 3\% regardless of the candidate brightness while for the RF RB classifier the FNR significantly degrades for fainter objects (from $\sim 5\%$ to $\sim 40\%$). \texttt{Braai} version $d6\_m7$, RF classifier version $t17\_f5\_c3$]{\includegraphics[width=0.47\textwidth]{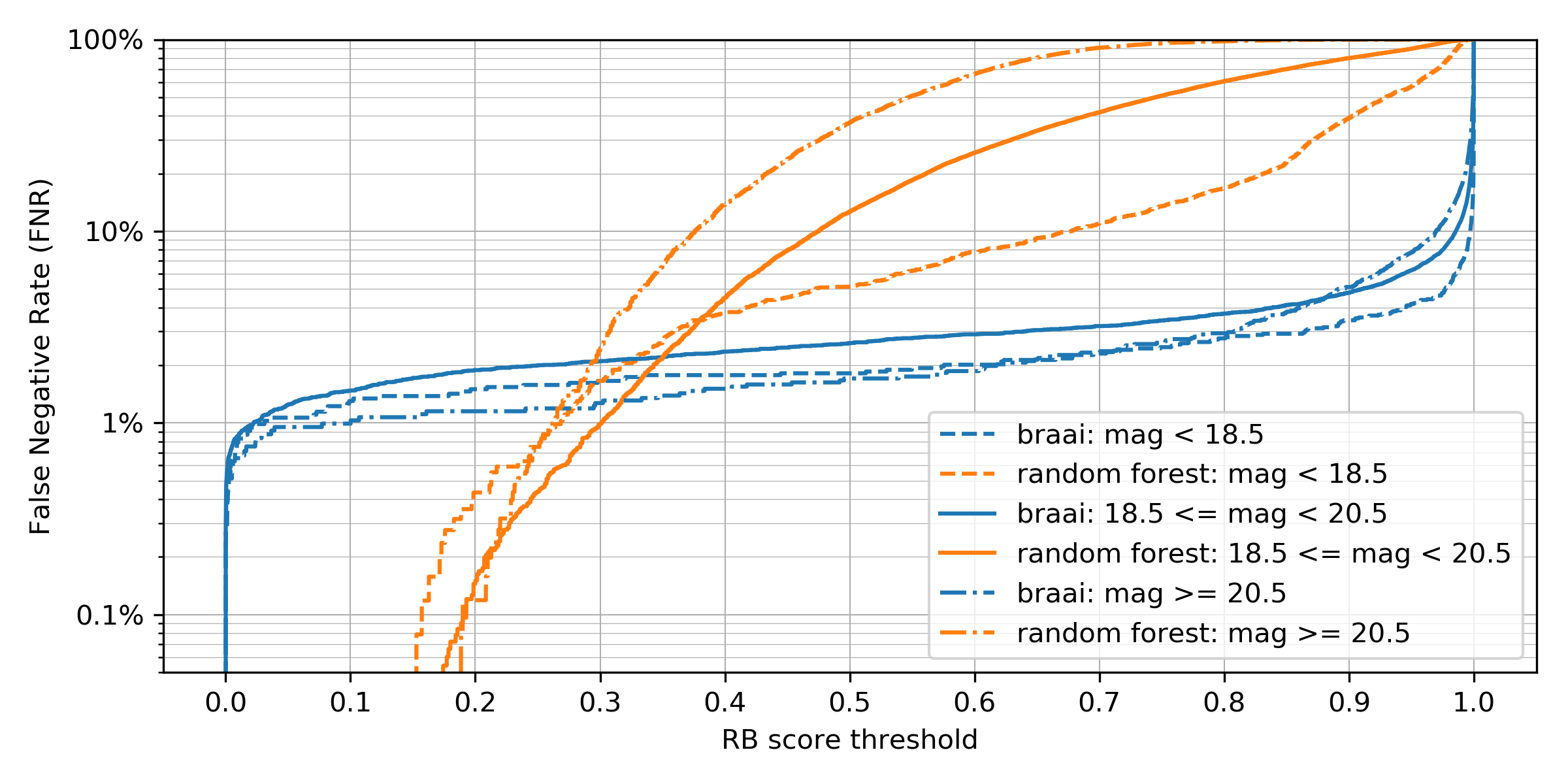}}
  \caption{\texttt{Braai} performance on a set of 21,641 alerts originating from known Solar system objects observed by ZTF in June 2019.}
    \label{fig:Asteroids}
\end{figure*}

Finally, we evaluated \texttt{braai} on 21,641 alerts originating from known Solar system objects\footnote{Non-streaking at the nominal 30-second ZTF exposure time.} observed by ZTF in June 2019. Since the candidates here are identified automatically, this set is not limited to high signal-to-noise levels that humans usually find easy to detect (see Fig. \ref{fig:Asteroids}a). Fig. \ref{fig:Asteroids}b shows the FNR of \texttt{braai} vs. RF RB classifier for asteroids with $Vmag < 18.5$ (2,534 alerts), $18.5 <= Vmag < 20.5$ (16,587 alerts), and $Vmag >= 20.5$ (2,520 alerts). At a score threshold of 0.5, \texttt{braai}'s FNR stays below 3\% regardless of the candidate brightness while for the RF RB classifier the FNR significantly degrades for fainter objects (from $\sim 5\%$ to $\sim 40\%$).

\subsection{Production deployment and Edge TPUs}

\texttt{Braai} was recently integrated into the ZSDS' image-differencing and event-extraction pipeline, which executes on a compute cluster of 66 commodity dual-socket Intel Xeon servers \citep{2019PASP..131a8003M}. The score and model version (\textit{drb} and \textit{drbversion}) are recorded in the \textit{candidate} block of each alert packet. The \texttt{braai} score is provided "as is" as a reliability metric and is not used to filter the outgoing alert stream. The individual science groups/users use different thresholds depending on the science case and their FNR/FPR requirements.

While, currently, the model, being relatively small in size, is evaluated on CPUs in production, we have experimented with alternative solutions. Concretely, we produced a version of \texttt{braai} that can be executed on Edge TPUs made by \texttt{Google} under the \texttt{Coral} brand -- a new class of efficient and cheap ($\sim\$100$) devices designed for heavy ML inference workloads.\footnote{See \url{https://coral.withgoogle.com/products/}} Currently, Edge TPUs can only operate with quantized models, i.e. both the model input and tensor parameters must be 8-bit fixed-point numbers. To achieve this, we performed quantization-aware training in \texttt{TensorFlow}, which uses ``fake'' quantization nodes to simulate the effect of 8-bit values during training, thus allowing inference to run using the quantized values. This technique makes the model more tolerant of the lower precision values, which generally results in a higher accuracy model (compared to post-training quantization)\footnote{\url{https://coral.withgoogle.com/docs/edgetpu/models-intro/}}. The quantized model is subsequently converted to \texttt{TensorFlow Lite} and compiled for Edge TPU usage.

This Edge TPU-native version of \texttt{braai} was deployed on a \texttt{Raspberry Pi 3 Model B+} single-board computer with a USB Edge TPU accelerator. Although the scores produced by it are scaled 8-bit integers and thus less numerically precise, \texttt{braai\_edgetpu} yields virtually the same performance as the full version of \texttt{braai} at a score threshold of 0.5 and it takes about ten minutes to process a typical night of ZTF alerts (200,000). We have also deployed \texttt{braai\_edgetpu} on the Edge-TPU-enabled \texttt{Coral} Dev board and were able to achieve a $4-5\times$ faster processing rate of up to 1200 triplets per second.\footnote{The main speed limiting factor for Raspberry Pi 3 Model B+ is that it only comes with a USB 2.0 interface.}

We stress that even quite complicated DL architectures can be efficiently executed on these devices with minimal effect on the inference accuracy making costly (potentially cloud-based) GPU work flows virtually unnecessary. This has significant implications for current and future surveys.

\section{Conclusions}

We have demonstrated that by putting together a large, representative, and uncontaminated data set with a relatively simple DL model we can achieve a state-of-the-art real/bogus classification performance. To improve it even further, we will retrain and deploy new classifiers as more labeled data are collected, especially in the $i$-band and at low Galactic latitudes, and if there are changes made to the hardware, or to intermediate readout and processing steps. With more data we may split the classifier for $g$, $r$, and $i$ bands, but the current performance suggests that that may not be required.


We note that the RB score provides only one reliability metric. Different ZTF science groups perform additional filtering using multiple alert columns (see \citet{Kasliwal_2019}).\footnote{GROWTH science program filters are mostly driven by particular scientific requirements.}

The data set that we put together for this project will be used in future work on DL system currently under development, such as specialized classifiers. 
It should be easy to reuse/extend this setup for other surveys including the Large Synoptic Survey Telescope (LSST, \citet{ivezic2008lsst}) in the not far future.

\vspace{\baselineskip}

\texttt{Braai} code and pre-trained models are available at \url{https://github.com/dmitryduev/braai}

\section*{Acknowledgements}
DAD and MJG acknowledge support from the Heising-Simons Foundation under Grant No. 12540303. AM and MJG acknowledge support from the NSF (1640818, AST-1815034), and IUSSTF (JC-001/2017). 
MMK acknowledges support by the GROWTH project funded by the NSF under Grant No. 1545949.
Based on observations obtained with the Samuel Oschin Telescope 48-inch Telescope at the Palomar Observatory as part of the Zwicky Transient Facility project. Major funding has been provided by the U.S. National Science Foundation under Grant No. AST-1440341 and by the ZTF partner institutions: the California Institute of Technology, the Oskar Klein Centre, the Weizmann Institute of Science, the University of Maryland, the University of Washington, Deutsches Elektronen-Synchrotron, the University of Wisconsin-Milwaukee, and the TANGO Program of the University System of Taiwan. Part of this research was carried out at the Jet Propulsion Laboratory, California Institute of Technology, under a contract with the National Aeronautics and Space Administration.

The authors are grateful to Eran Ofek for useful discussions.




\bibliographystyle{mnras}
\bibliography{braai}







\bsp	
\label{lastpage}
\end{document}